\documentclass[11pt]{scrartcl}

\usepackage[utf8]{inputenc} 
\usepackage{mathtools, amsfonts, amssymb, breqn}
\usepackage{amsthm}

\usepackage[british]{babel}
\usepackage{graphicx, hyperref, tabularx, abstract, makecell,ragged2e}

\usepackage[format=plain]{caption}
\DeclareCaptionFormat{cont}{#1 (cont.)#2#3\par}

\usepackage[outdir=./]{epstopdf}   

\usepackage{cellspace, hhline} 

\usepackage{subfig}
\usepackage{xcolor-solarized}
\usepackage{todonotes}
\usepackage{overpic}

\usepackage{enumitem}
\setdescription{itemsep=5pt,parsep=0pt,leftmargin=1.55cm,font=\normalfont}

\usepackage[]{geometry}
\newcommand{\N}{\ensuremath{\mathbb{N}}}
\newcommand{\R}{\ensuremath{\mathbb{R}}}
\newcommand{\C}{\ensuremath{\mathbb{C}}}

\newcommand{\mattwo}[4]{\left(
	\begin{array}{cc}
		#1 & #2 \\
		#3 & #4 \\  
	\end{array}
	\right)
}
\newcommand{\matthree}[9]{\left(
	\begin{array}{ccc}
		#1 & #2 & #3 \\
		#4 & #5 & #6 \\  
		#7 & #8 & #9 
	\end{array}
	\right)
}

\newcommand{\tr}{{\operatorname{tr}}}

\usepackage{cellspace, hhline}

\numberwithin{equation}{section}
\numberwithin{figure}{section}
\numberwithin{table}{section}

\begin{document}
	\author{L. Eigentler and J. A. Sherratt}
	
	\newcommand{\Addresses}{{
			\bigskip
			\footnotesize
			
			L. Eigentler (corresponding author),  Maxwell Institute for Mathematical Sciences, Department of Mathematics, Heriot-Watt University, Edinburgh EH14 4AS, United Kingdom\par\nopagebreak
			\textit{E-mail address}: \texttt{le8@hw.ac.uk}
			
			\medskip
			
			J. A. Sherratt,  Maxwell Institute for Mathematical Sciences, Department of Mathematics, Heriot-Watt University, Edinburgh EH14 4AS, United Kingdom\par\nopagebreak
			\textit{E-mail address}: \texttt{J.A.Sherratt@hw.ac.uk}

	}}
	
	\title{Metastability as a coexistence mechanism in a model for dryland vegetation patterns}
	\date{}
	\maketitle
	\Addresses

\begin{abstract}
Vegetation patterns are a ubiquitous feature of water-deprived ecosystems. Despite the competition for the same limiting resource, coexistence of several plant species is commonly observed. We propose a two-species reaction-diffusion model based on the single-species Klausmeier model, to analytically investigate the existence of states in which both species coexist. Ecologically, the study finds that coexistence is supported if there is a small difference in the plant species' average fitness, measured by the ratio of a species' capabilities to convert water into new biomass to its mortality rate. Mathematically, coexistence is not a stable solution of the system, but both spatially uniform and patterned coexistence states occur as metastable states. In this context, a metastable solution in which both species coexist corresponds to a long transient (exceeding $10^3$ years in dimensional parameters) to a stable one-species state. This behaviour is characterised by the small size of a positive eigenvalue which has the same order of magnitude as the average fitness difference between the two species. Two mechanisms causing the occurrence of metastable solutions are established: a spatially uniform unstable equilibrium and a stable one-species pattern which is unstable to the introduction of a competitor. We further discuss effects of asymmetric interspecific competition (e.g. shading) on the metastability property.

\end{abstract}


\section{Introduction}

Vegetation patterns in semi-arid climate zones are a prime example of a self-organising principle in ecology \cite{Valentin1999, Deblauwe2008}. One of the main mechanisms that creates such a mosaic of biomass and bare soil is a modification of soil properties by plants that induces a water redistribution feedback loop \cite{Rietkerk2008,Meron2012,Meron2016,Meron2018}. On bare ground only small amounts of water are able to infiltrate into the soil and water run-off occurs, while in regions covered by biomass the soil's water infiltration capacity is increased. Dense plant patches therefore act as sinks and deplete soil water in regions of bare ground \cite{Thompson2010,Eldridge2000}. This redistribution of the limiting resource drives further growth in vegetation patches and thus closes the feedback loop.

Drylands account for approximately 41\% of the Earth's land mass and are home to a similar proportion (38\%) of the world's human population. The sizes of arid and semi-arid regions that suffer from land degradation are expected to increase over the coming decades due to climate change \cite{Reynolds2007}. Vegetation patterns are a characteristic feature of such fragile ecosystems. Patterns have been detected in semi-desert regions in the African Sahel \cite{Thiery1995,White1971,Worrall1959, Mueller2013,Deblauwe2012}, Somalia \cite{Hemming1965,Gowda2018}, Australia \cite{Dunkerley2002,Tongway1990,Heras2012}, Israel \cite{Sheffer2013,Buis2009} and Mexico and the US \cite{Cornet1988,Montana1990,Montana1992,Deblauwe2012, Penny2013,Pelletier2012}. Changes to characteristic features of vegetation patterns in these regions such as the pattern wavelength, the area fraction covered by biomass, or the recovery time from perturbations can act as early indicators of desertification as they provide a useful tool in predicting further changes to ecosystems \cite{Dakos2011,Saco2018,Rietkerk2004,Kefi2007,Gowda2016,Corrado2014,Zelnik2018}. This is an issue of considerable socio-economic importance since agriculture is a major contributor to the economy in many drylands \cite{UNGlobalLandOutlook2017}. For example, in sub-Saharan Africa livestock frequently graze on spatially patterned vegetation. Thus changes in vegetation levels have a major effect on the livestock sector, which makes a very significant contribution to GDP, e.g.\ 20\% in Chad, 15\% in Mali, 12\% in Niger and 7.5\% in Burkina Faso \cite{UNLivestockBrief2005,Dickovick2014}, with involvement of high proportions of the population (e.g.\ 40\% in Chad \cite{Dickovick2014}). 

Due to the temporal and spatial scales involved in the evolution of vegetation patterns, these ecosystems cannot be recreated in a laboratory setting. To gain a better understanding of the pattern dynamics a number of mathematical models have been proposed (see \cite{Borgogno2009, Zelnik2013} for reviews). In particular, modelling efforts based on partial differential equations, most notably by \cite{Gilad2004,Gilad2007} and \cite{HilleRisLambers2001,Rietkerk2002}, provide a rich framework for mathematical analysis. One model that stands out due to its simplicity is the Klausmeier model \cite{Klausmeier1999}, which provides a deliberately basic description of the plant-water dynamics in semi-arid environments. The highly accessible nature of the model enables a detailed model analysis (e.g. by \cite{Sherratt2005,Sherratt2007,Sherratt2010,Sherratt2011,Sherratt2013III,Sherratt2013IV, Sherratt2013V,Ursino2006,Siteur2014,Eigentler2018nonlocalKlausmeier,Bennett2019, Consolo2019}). Recent advances in remote sensing technology using satellite data provide a promising tool to test model predictions on pattern resilience \cite{Bastiaansen2018,Gandhi2018}.

Most models in this context only consider a single plant species or combine several species into one single variable. However, vegetation patches often consist of a mix of herbaceous and woody species, where the latter can usually be found in the centre of a patch, surrounded by the former \cite{Herbes2001,Seghieri1997}. Previous simulation-based studies of dryland ecosystem models have indeed been able to reproduce patterns in which two species coexist by considering a variety of different mechanisms and feedbacks that enable diversity in ecosystems \cite{Gilad2007a,Nathan2013,Ursino2016,Callegaro2018,Baudena2013}. One such facilitative mechanism occurs in a system of two species in which only one plant type induces a pattern forming feedback. If, additionally, the non-pattern forming species is superior in its water uptake and dispersal capabilities, then the pattern-forming species can act as an ecosystem engineer to facilitate coexistence of both species in patterned form \cite{Nathan2013,Baudena2013}. Even if patterns in which two species coexist are not observed as long-term solutions of a system, they can feature in a transition between two stable states. \cite{Gilad2007a} briefly report on the observation of coexistence patterns as a slow (several hundred years) transient during which patterns form due to facilitation between two species before eventually one of the species becomes extinct as competitive feedbacks take over. A different mechanism that enables coexistence of two species in both uniform and spatially patterned settings is adaptation to different ecological niches, such as soil moisture \cite{Ursino2006,Callegaro2018}.

In-phase spatial patterns are not the only phenomenon that is studied in the context of species coexistence. The existence of a multitude of localised patterns of one species in an otherwise uniform cover of a second species (homoclinic snaking) has also been observed as a possible form of coexistence in a mathematical model \cite{Kyriazopoulos2014}.  The solution arises from a model that assumes a trade-off between root and shoot growth causing a balance between the competition for water and for light that supports coexistence. Other models are not able to make any statement on the coexistence of species, but yield valuable information on facilitation and competition between the plant types based on differences in traits such as their dispersal behaviour \cite{Pueyo2010}.

The savanna biome has also been studied by various non-spatial models that describe the dynamics of the relative abundances of grass, trees and water. While such models are unable to make any statements on the formation of spatial patterns, they still provide valuable insights into coexistence-preserving effects of processes such as precipitation intermittency \cite{DOnofrio2015}, facilitation by grasses towards trees \cite{Synodinos2015} or fire disturbances \cite{Baudena2010,Scheiter2007}.

Previous model analysis on species coexistence in semi-arid landscapes has  mainly focussed on feedback loops induced through differences in the plant species' traits and their effects on multi-species plant communities. We are not aware of any studies that investigate effects of the differences in basic properties such as plant mortality or plant growth rate on semi-arid vegetation patterns. In this paper we aim to analytically address the question how the difference between two plant types can give rise to a multispecies metastable vegetation pattern (a unstable pattern whose instability is caused by a very small unstable eigenvalue \cite{Potapov2005}) and how the pattern's properties are affected by changes to the difference between the species. 

To do this we introduce a multi-species model based on the Klausmeier model in Section \ref{sec: Multispecies: Model}. Numerical simulations of the model presented in Section \ref{sec: Multispecies: Simulations} suggest two different origins of metastable coexistence patterns. These two pathways into the problem are closely examined through a linear stability analysis in Sections \ref{sec: Multispecies: Metastable patterns from tree patterns} and \ref{sec: Multispecies: Metastable patterns from coexistence steady state}. Finally, we discuss our results in Section \ref{sec: Multispecies: Discussion}.


\section{Model}\label{sec: Multispecies: Model}
In this section we lay out the framework used in this paper to analyse the coexistence of grass and trees in dryland ecosystems. We propose a model based on the extended Klausmeier model \cite{Klausmeier1999}, which in dimensional form is
\begin{subequations}\label{eq: Multispecies: Model: Klausmeier}
	\begin{align}\label{eq: Multispecies: Model: Klausmeier plants}
	\frac{\partial u}{\partial t} &= \overbrace{c_1c_2u^2w}^{\text{plant growth}} -\overbrace{ c_3 u}^{\substack{\text{plant}\\\text{mortality}}} +\overbrace{ c_4\frac{\partial^2 u}{\partial x^2}}^{\text{plant dispersal}}, \\
	\frac{\partial w}{\partial t} &= \underbrace{c_5}_{\text{rainfall}}-\underbrace{c_6w}_{\text{evaporation}} - \underbrace{c_2u^2w}_{\substack{\text{water uptake}\\\text{by plants}}} + \underbrace{c_7 \frac{\partial w}{\partial x}}_{\substack{\text{water}\\\text{advection}}} + \underbrace{c_8 \frac{\partial^2 w}{\partial x^2}}_{\substack{\text{water}\\\text{diffusion}}},
	\end{align}
\end{subequations}
where $u(x,t)$ is the weight of plants per unit area and $w(x,t)$ is the mass of water per unit area in the one-dimensional space domain $x\in\R$ at time $t>0$. The water supply (precipitation) of the system is assumed to be constant at rate $c_5$, while evaporation and plant loss effects are assumed to be proportional to the respective densities at rates $c_6$ and $c_3$, respectively. The nonlinearity in the terms describing water uptake and biomass growth arises due to a soil modification by plants. The term is the product of the density of the consumer $u$ and of the available resource $c_2uw$, which corresponds to water being able to infiltrate into the soil. The dependence on the plant density $u$ in the latter term occurs due to a positive correlation between the plant density and the soil surface's permeability \cite{Rietkerk2000, Valentin1999, Cornet1988}. Plant growth is assumed to be proportional to water uptake \cite{Rodriguez-Iturbe1999, Salvucci2001} and water to biomass conversion takes place at rate $c_1$. In its original setting, the Klausmeier model is formulated to describe the dynamics on sloped terrain on which water flow downhill is modelled by advection at rate $c_7$. An extension includes diffusion of water at rate $c_8$ to account for water redistribution on flat ground and is well established now (e.g. \cite{Siteur2014,Stelt2013,Zelnik2013,Kealy2012}). Plant dispersal is also modelled by a diffusion term (with diffusion rate $c_4$).

Both on flat ground and on sloped terrain \eqref{eq: Multispecies: Model: Klausmeier} captures the formation of patterns for sufficiently low levels of precipitation and their properties have been studied extensively \cite{Klausmeier1999, Sherratt2005, Sherratt2007, Sherratt2010, Sherratt2011, Sherratt2013III, Sherratt2013IV, Sherratt2013V, Siteur2014}. In \eqref{eq: Multispecies: Model: Klausmeier} the plant density $u$ either describes one single species or accounts for the totality of all plant types in the ecosystem. While an ecosystem rarely consists of only one single species, estimation of species-dependent parameters such as the plant mortality rate $c_3$ may be impractical if $u$ is comprised of many different species for which parameter estimates differ significantly (see for example estimates for tree and grass species by \cite{Klausmeier1999}). 

An extension of \eqref{eq: Multispecies: Model: Klausmeier} that accounts for the differences between plant species in the same ecosystem can be obtained by separating the plant density $u$ into $n\in\N$ different species $u_i$, $i=1,\dots, n$ that satisfy \eqref{eq: Multispecies: Model: Klausmeier} with an appropriate set of parameters in the absence of all other species. The model arising from this assumption is

\begin{subequations}\label{eq: Multispecies: Model: general multispecies no shading}
	\begin{align}
	\frac{\partial u_i}{\partial t} &= \overbrace{k_1^{(i)}wu_i \left(\sum_{j=1}^n k_2^{(j)}u_j\right)}^{\text{plant growth}} - \overbrace{k_3^{(i)} u_i}^{\substack{\text{plant}\\\text{mortality}}}+ \overbrace{k_5^{(i)}\frac{\partial^2 u_i}{\partial x^2}}^{\text{plant dispersal}}, \\
	\frac{\partial w}{\partial t} &= \underbrace{k_6}_{\text{rainfall}}-\underbrace{k_7w}_{\text{evaporation}} - \underbrace{w \left(\sum_{j=1}^n u_j\right) \left( \sum_{j=1}^n k_2^{(j)}u_j\right)}_{\text{water uptake by plants}} + \underbrace{k_8 \frac{\partial w}{\partial x}}_{\substack{\text{water}\\\text{advection}}}  + \underbrace{k_9 \frac{\partial^2 w}{\partial x^2}}_{\substack{\text{water}\\\text{diffusion}}}.
	\end{align}
\end{subequations}
for $i=1, \dots, n$. In this multi-species model, the term describing water uptake by plants is, as in \eqref{eq: Multispecies: Model: Klausmeier}, the product of the water density $w$, the total plant density $\sum_{j=1}^{n} u_j$ and the soil's infiltration capacity  $\sum_{j=1}^{n} k_2^{(j)} u_j$. The species-dependent constants $k_2^{(i)}$ account for the plant types' different contributions to the soil properties. The summands in $\sum_{j=1}^{n} u_j$ correspond to the consumption of water by each single species and are therefore not replicated in the term describing plant growth. Thus, the addition of new biomass of species $u_i$ with water to biomass conversion rate $k_1^{(i)}$ only depends on the water density, the soil's infiltration capacity and the density of species $u_i$ itself. The remaining assumptions are identical to those taken in the formulation of \eqref{eq: Multispecies: Model: Klausmeier}, i.e. $k_3^{(i)}$ and $k_5^{(i)}$ denote the mortality and diffusion rates of species $u_i$, respectively; $k_6$ is the constant amount of rainfall which adds water to the system; and $k_7$, $k_8$, and $k_9$ are the evaporation, advection and diffusion rates of water, respectively.

In \eqref{eq: Multispecies: Model: general multispecies no shading} no direct interspecific interaction takes place. Instead plant species only compete indirectly through depletion of the limiting resource - water. Models of this type, in which species compete for the same limiting resource without any direct competition between the different types, do not provide a framework able to describe coexistence as the species that can tolerate the lowest level of the limiting resource outcompetes all competitors \cite{Tilman1982}. Thus, a description of an ecosystem in which plant species coexist needs to take interspecific dynamics, such as shading, into account.

For simplicity we restrict the model to a system on flat ground of two plant species $u_1$ and $u_2$ only, in which one species inhibits the other by increasing its competitor's mortality rate but its own mortality rate remains unaffected by the presence of the other species. An alternative approach to model direct interspecific competition would be a reduction of a species' biomass growth rate \cite{Kyriazopoulos2014}. A classic example of such an one-sided inhibitory direct interaction is two species, such as a herbaceous and a woody species, where the latter grows much taller than the former and thus imposes a shading effect on its competitor. Shading may also have a facilitative effect on plants and induce a positive feedback loop due to a reduction in evaporation \cite{Gilad2007a, Baudena2013}. In contrast to a one-sided inhibitory shading effect, shading-induced evaporation reduction affects both species as beneficial effects occur indirectly through a variation in resource availability. Thus, the nonlinearity in the plant densities of the water consumption and plant growth terms can account for such a beneficial effect as it collectively describes all positive feedback loops increasing the growth of biomass.

Adding an inhibitory shading term to \eqref{eq: Multispecies: Model: general multispecies no shading} with $n=2$, we propose the model studied in this paper, which is
\begin{subequations}\label{eq: Multispecies: Model: dimensional model}
	\begin{align}
	\frac{\partial u_1}{\partial t} &= \overbrace{k_1^{(1)}wu_1\left(k_2^{(1)}u_1 + k_2^{(2)}u_2\right)}^{\text{plant growth}} - \overbrace{k_3^{(1)} u_1}^{\substack{\text{plant}\\\text{mortality}}}-\overbrace{k_4u_1u_2}^{\substack{\text{interspecific}\\\text{competition}}} + \overbrace{k_5^{(1)}\frac{\partial^2 u_1}{\partial x^2}}^{\text{plant dispersal}}, \\
	\frac{\partial u_2}{\partial t} &= \overbrace{k_1^{(2)}wu_2\left(k_2^{(1)}u_1 + k_2^{(2)}u_2\right)}^{\text{plant growth}} -\overbrace{ k_3^{(2)} u_2}^{\substack{\text{plant}\\\text{mortality}}} +\overbrace{ k_5^{(2)}\frac{\partial^2 u_2}{\partial x^2}}^{\text{plant dispersal}}, \\
	\frac{\partial w}{\partial t} &= \underbrace{k_6}_{\text{rainfall}}-\underbrace{k_7w}_{\text{evaporation}} - \underbrace{w\left(u_1+u_2\right)\left(k_2^{(1)}u_1 + k_2^{(2)}u_2\right)}_{\text{water uptake by plants}} + \underbrace{k_9 \frac{\partial^2 w}{\partial x^2}}_{\substack{\text{water}\\\text{diffusion}}}.
	\end{align}
\end{subequations}
The shading effect causes species $u_2$ to impose an additional mortality effect on $u_1$ that is dependent on the density $u_2$, while $u_1$ does not have such an effect on $u_2$. The results presented in this paper are robust to changes in the functional response of this shading effect. Results for shading effects with a Holling type II and Holling type III functional response show no qualitative difference to the algebraically simpler term in \eqref{eq: Multispecies: Model: dimensional model}. Table \ref{tab: Multispecies: Models: Parameters} provides an overview of parameter estimates used in the model. As indicated in the table, we were able to obtain estimates for parameters from previous models on dryland vegetation, except for the rate of the direct interspecific interaction $k_4$. However, our model analysis in Sections \ref{sec: Multispecies: Metastable patterns from tree patterns} and \ref{sec: Multispecies: Metastable patterns from coexistence steady state} suggests a suitable range for the shading parameter that yields biologically relevant results and we briefly discuss effects caused by deviations from this range.

\begin{table}
	\begin{tabularx}{\textwidth}{lllX}
		\hline
		Parameter & Units & Estimates & Description \\
		\hline
		$k_1^{(1)}$ &\makecell[l]{ (kg biomass) \\(kg H$_2$0)$^{-1}$} 	& \makecell[l]{0.003 [1], \\ 0.007 [2]} & Water to biomass conversion rate for species $u_1$ \\
		$k_1^{(2)}$ & \makecell[l]{(kg biomass)\\ (kg H$_2$0)$^{-1}$} 	& \makecell[l]{0.002 [1], \\ 0-0.01 [2]} & Water to biomass conversion rate for species $u_2$ \\
		$k_2^{(1)}$ &\makecell[l]{m$^4$ year$^{-1}$\\ (kg biomass)$^{-2}$} & \makecell[l] {100 [1]} & Effect of plant species $u_1$ on water infiltration into the soil \\
		$k_2^{(2)}$ & \makecell[l]{m$^4$ year$^{-1}$\\ (kg biomass)$^{-2}$} & \makecell[l] {1.5 [1]} & Effect of plant species $u_2$ on water infiltration into the soil \\
		$k_3^{(1)}$ & year$^{-1}$ 						& \makecell[l]{1 [3], \\ 1.8 [1]}	& Rate of plant loss for species $u_1$\\
		$k_3^{(2)}$ & year$^{-1}$ 						& \makecell[l]{0.023 [3] \\ 0.18 [1], 1.2 [4]}	& Rate of plant loss for species $u_2$ \\
		$k_4$ & \makecell[l]{m$^2$ year$^{-1}$\\ (kg biomass)$^{-1}$}  & - (see text) & Interspecific competition (shading) \\
		$k_5^{(1)}$ & m$^2$ year$^{-1}$ 					& 1 [1,5], 36.5 [2] & Rate of diffusion of $u_1$ \\
		$k_5^{(2)}$ & m$^2$ year$^{-1}$ 					& \makecell[l]{$6.25 \cdot 10^{-4}$ [4], \\ 1 [1]} & Rate of diffusion of $u_2$ \\
		$k_6$ & \makecell[l]{(kg H$_2$0) \\ m$^{-2}$ year$^{-1}$}	& \makecell[l]{250-750 [1],\\ 0-1000 [4], \\ 150-1200 [3],\\ 0-365 [2]} & Rainfall \\
		$k_7$ & year$^{-1}$ 						& \makecell[l]{2 [2], 4 [1,4,5], \\ 8 [3] } & Rate of evaporation \\
		$k_9$ & m$^2$ year$^{-1}$  & 500 [5], & Rate of water diffusion \\ 	\hline	
		Parameter & Scaling & Estimates & Description \\
		\hline
		$A$ & $k_1^{(1)} (k_2^{(1)} )^{\frac{1}{2}} k_6 k_7^{-\frac{3}{2}}$ 									& 0.94-2.8 [1] & Nondimensionalised constant corresponding to rainfall \\
		$B_1$ & $k_3^{(1)}k_7^{-1}$									 	& \makecell[l]{0.125 [3], 0.45 [1]} & Nondimensionalised constant corresponding to the rate of plant loss of $u_1$\\
		$B_2$ & $k_3^{(2)}k_7^{-1}$									 	& \makecell[l]{0.0029 [3], \\ 0.045 [1]} & Nondimensionalised constant corresponding to the rate of plant loss of $u_2$\\
		$F$ & $k_1^{(2)}(k_1^{(1)})^{-1}$ 									& 0-1 [2], 0.67 [1] & Ratio of plants' water to biomass conversion rates \\
		$H$ & $k_2^{(2)} (k_2^{(1)})^{-1}$ 									& 0.015 [1] & Ratio of plants' effects on water infiltration into soil  \\
		$S$ & $k_4(k_2^{(1)}k_7)^{-\frac{1}{2}}$ & - (see text) & Nondimensional constant corresponding to shading effect \\
		$D$ & $k_5^{(2)} (k_5^{(1)} )^{-1}$									& 0-1 [1,2,4] &  Ratio of plant species' diffusion rates \\
		$d$ & $k_9 (k_5^{(1)} )^{-1}$									& 500 [5] &  Ratio of water and plant species $u_1$ diffusion rates
	\end{tabularx}
	\caption{Overview of parameters in \eqref{eq: Multispecies: Model: dimensional model} and \eqref{eq: Multispecies: Model: nondimensional model}. This table shows both the dimensional parameters in model \eqref{eq: Multispecies: Model: dimensional model} and the nondimensional parameters in \eqref{eq: Multispecies: Model: nondimensional model}, including their units (dimensional parameters) or scalings (nondimensional parameters), the estimated values that we use, and a brief description. The parameter estimates are obtained from [1]: \cite{Klausmeier1999}, [2]: \cite{Baudena2013}, [3]: \cite{Synodinos2015}, [4]: \cite{Gilad2007} and [5]: \cite{Siteur2014}.}\label{tab: Multispecies: Models: Parameters}
\end{table}

A suitable nondimensionalisation for the model is 
\begin{align*}
u_1&=\left(\frac{k_7}{k_2^{(1)}}\right)^{\frac{1}{2}}\widetilde{u_1}, \quad u_2=\left(\frac{k_7}{k_2^{(1)}}\right)^{\frac{1}{2}}\widetilde{u_2}, \quad w=\frac{k_7^{\frac{1}{2}}}{k_1^{(1)}\left(k_2^{(1)} \right)^{\frac{1}{2}}} \widetilde{w}, \\ x &= \left(\frac{k_5^{(1)}}{k_7} \right)^{\frac{1}{2}} \widetilde{x}, \quad t = \frac{1}{k_7} \widetilde{t}.
\end{align*}
The model thus becomes
\begin{subequations}\label{eq: Multispecies: Model: nondimensional model}
	\begin{align}
	\frac{\partial u_1}{\partial t} &= wu_1\left(u_1 + Hu_2\right) - B_1 u_1-Su_1u_2 + \frac{\partial^2 u_1}{\partial x^2},\label{eq: Multispecies: Model: nondimensional model u1} \\
	\frac{\partial u_2}{\partial t} &=Fwu_2\left(u_1 + Hu_2\right) - B_2 u_2 +D\frac{\partial^2 u_2}{\partial x^2}, \label{eq: Multispecies: Model: nondimensional model u2}\\
	\frac{\partial w}{\partial t} &= A-w - w\left(u_1+u_2\right)\left(u_1 + Hu_2\right) +d \frac{\partial^2 w}{\partial x^2},
	\end{align}
\end{subequations}
after dropping the $\widetilde{\cdot}$'s for brevity, where
\begin{align*}
&A = \frac{k_1^{(1)} \left(k_2^{(1)} \right)^{\frac{1}{2}} k_6}{k_7^{\frac{3}{2}}}, \quad B_1 = \frac{k_3^{(1)}}{k_7}, \quad B_2 = \frac{k_3^{(2)}}{k_7}, \quad S=\frac{k_4}{\left(k_2^{(1)}k_7\right)^{\frac{1}{2}}}, \\ &F = \frac{k_1^{(2)}}{k_1^{(1)}}, \quad H = \frac{k_2^{(2)}}{k_2^{(1)}}, \quad D = \frac{k_5^{(2)}}{k_5^{(1)}}, \quad d=\frac{k_9}{k_5^{(1)}}.
\end{align*}
The constants $A$ and $B_i$ are combinations of several of the original model's parameters, but represent rainfall and plant mortality, respectively. The ratios $F$, $H$ and $D$ describe the differences in the plant species' water to biomass conversion rates, the effects on the soil's infiltration capacity and the diffusion coefficients, respectively. Finally, $d$ quantifies the ratio of the rate of water diffusion to that of the diffusion of plant species $u_1$. Table \ref{tab: Multispecies: Models: Parameters} includes estimates for the nondimensional parameters. 

In the analysis of the model we assume that $u_1$ is a herbaceous species and allow $u_2$ to vary between another grass species and a woody vegetation type. The parameters of $u_1$ are fixed throughout the analysis and act as a reference point. To investigate how the difference between two plant species affects the plant-water dynamics of the system, the parameters of $u_2$ are varied and comparisons to the fixed species $u_1$ are made. For brevity we refer to the two plant densities as grass and trees, even if $u_2$ differs only slightly from $u_1$. Parameter estimates (see Table \ref{tab: Multispecies: Models: Parameters}) suggest that trees' rate of mortality is less than that of grasses ($B_2<B_1$), trees convert water into biomass less efficiently ($F<1$), trees affect the soil's water infiltration rate less severely per unit biomass ($H<1$) and trees disperse at a slower rate than grass ($D<1$). We further assume that the inhibitory effect of shading intensifies as the species difference increases. Thus, only this parameter region is analysed. In particular, to define a measure of species difference, we introduce a parameter $\chi \in [0,1]$ that describes the extent to which the species differ. Thus, we set

\begin{align} \label{eq: Multispecies: Model: parameter region to compare species}
\begin{split}
B_2 &= B_1-\chi(B_1-b_2), \quad F=1-\chi(1-f), \quad H=1-\chi(1-h), \\  S &=s\chi, \quad D=1-\chi(1-D_0),
\end{split} 
\end{align}

where $B_1$ is set to a typical mortality rate of a herbaceous species, $b_2$ to that of a woody species and $f$, $h$ and $D_0$ to the smallest respective ratios between two differing species. If the species are the same (i.e. $\chi=0$), then $B_2=B_1$, $F=H=D=1$ and $S=0$. In this case, \eqref{eq: Multispecies: Model: nondimensional model} simplifies to
\begin{subequations}\label{eq: Multispecies: Model: species same model}
	\begin{align}
	\frac{\partial \left(u_1+u_2\right)}{\partial t} &= w\left(u_1+u_2\right)^2 - B_1 \left(u_1+u_2\right) + \frac{\partial^2 \left(u_1+u_2\right)}{\partial x^2}, \\
	\frac{\partial w}{\partial t} &= A-w - w\left(u_1+u_2\right)^2 +d \frac{\partial^2 w}{\partial x^2},
	\end{align}
\end{subequations}
by adding \eqref{eq: Multispecies: Model: nondimensional model u1} and \eqref{eq: Multispecies: Model: nondimensional model u2}. This simplified model is the extended Klausmeier model \eqref{eq: Multispecies: Model: Klausmeier} in nondimensional form on flat ground for plant density $u_1+u_2$ and water density $w$.


\section{Numerical Solutions of the Model}\label{sec: Multispecies: Simulations}
To motivate the analysis presented in Sections \ref{sec: Multispecies: Metastable patterns from tree patterns} and \ref{sec: Multispecies: Metastable patterns from coexistence steady state} we present some typical solutions of \eqref{eq: Multispecies: Model: nondimensional model} that are obtained by numerical integration. Despite the inclusion of direct interspecific competition in \eqref{eq: Multispecies: Model: nondimensional model} and the associated existence of a pair of equilibria in which both species coexist (see Section \ref{sec: Multispecies: Metastable patterns from coexistence steady state}), the system converges to a single-species state for any choice of parameters. The nature of this long-term behaviour depends on the parameter values used in the integration and may be a uniform or patterned state of either species. The transient to such an equilibrium state in which only one of the plant types is present may, however, occur as a very slow process (exceeding $10^3$ years in dimensional parameters) in which both species coexist in either a patterned configuration or uniformly in space. Such a unstable state which nevertheless persists as a solution for a very long time (compared to the time taken to emerge from some initial configuration) is referred to as a metastable state in this context.

\begin{figure}
	\centering
	\subfloat[\label{fig: Multispecies: Simulations: example solutions coex patterns initally coex ss}]{\includegraphics[width=0.5\textwidth]{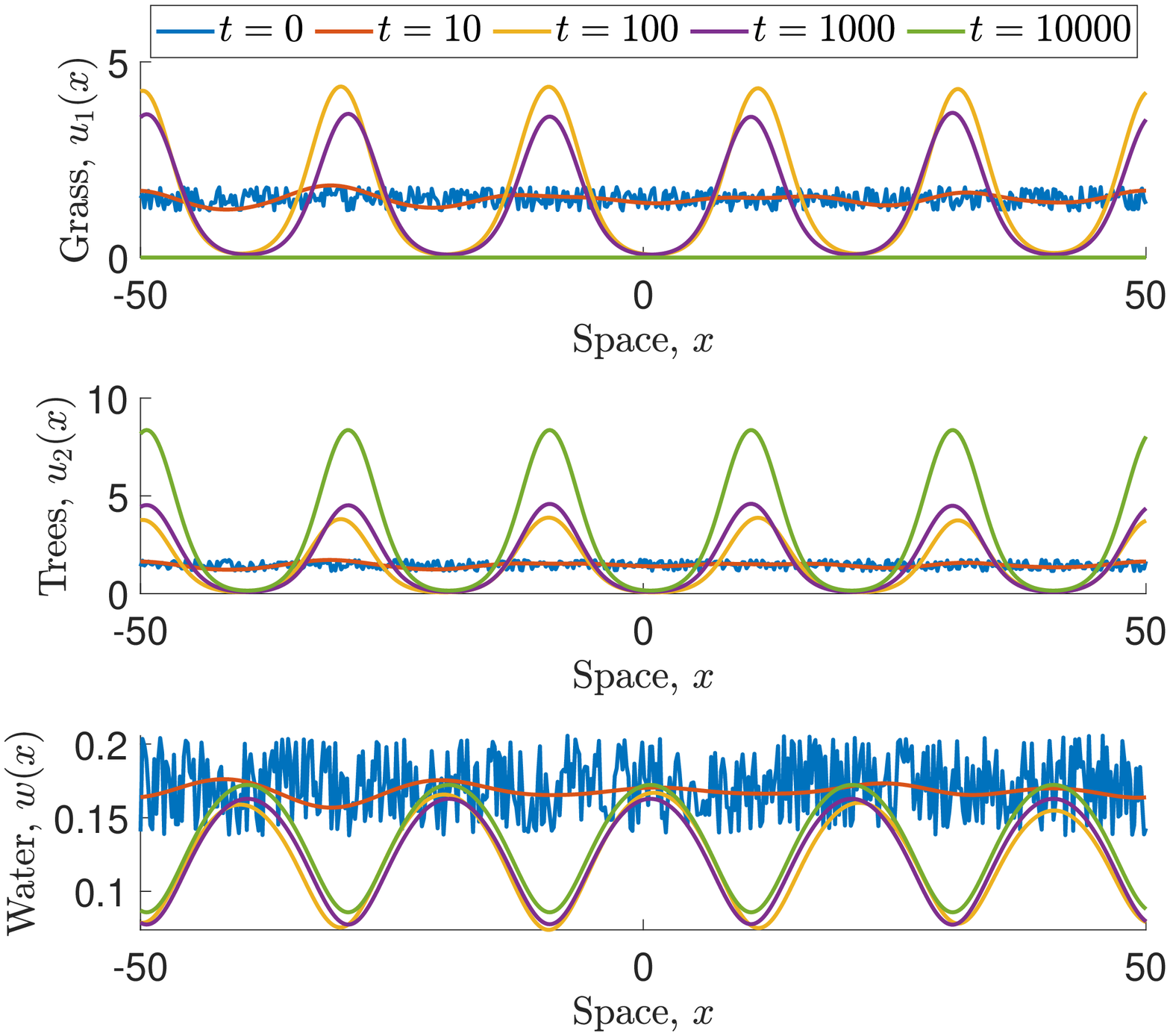}}
	\subfloat[\label{fig: Multispecies: Simulations: example solutions coex patterns initially tree pattern}]{\includegraphics[width=0.5\textwidth]{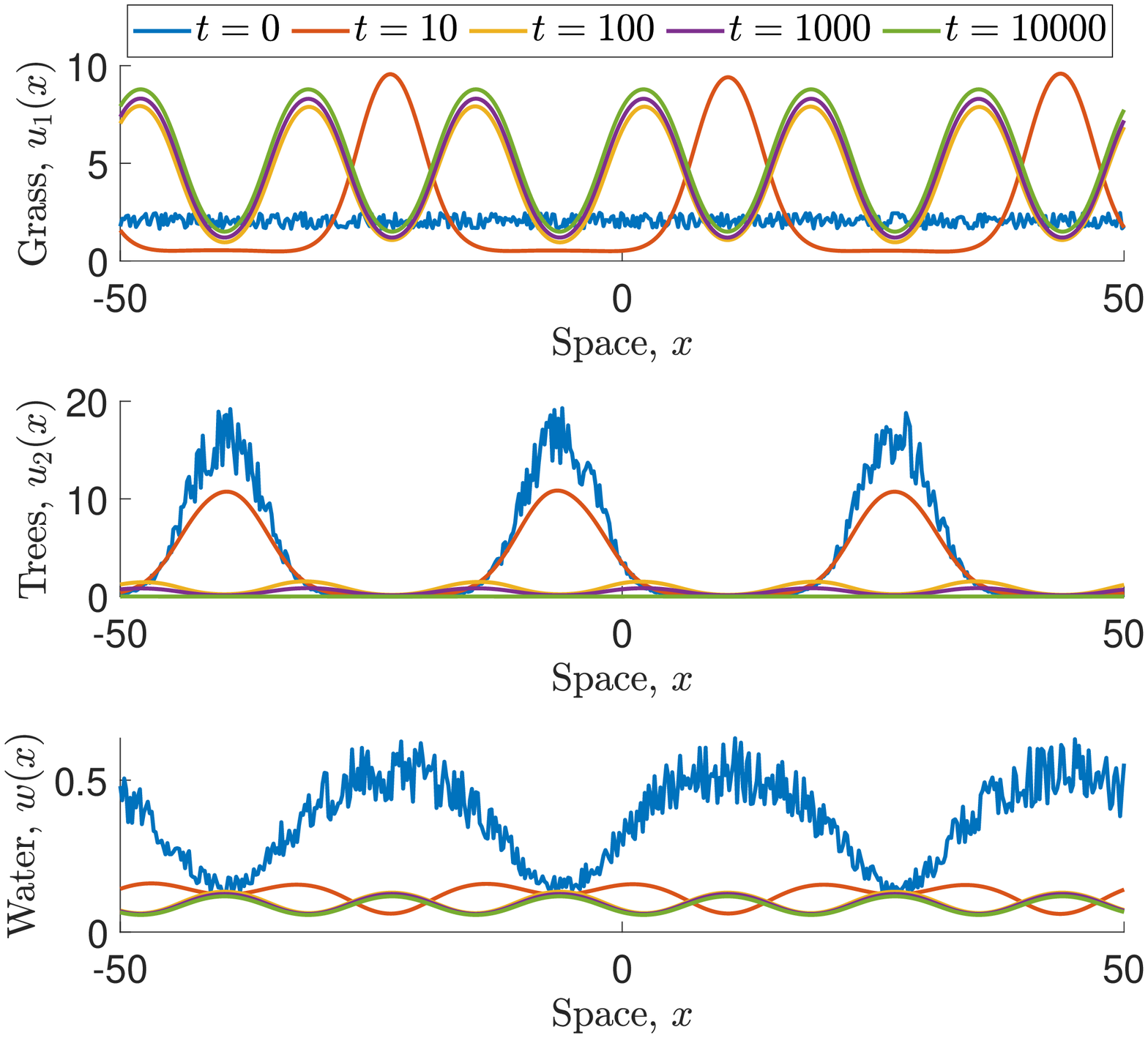}}
	\caption{Numerical solution of the multi-species model \eqref{eq: Multispecies: Model: nondimensional model} showing metastable patterns of species coexistence. The simulations are performed by discretising the space domain into $M\in \N$ equidistant points, which yields a system of $3M$ ordinary differential equations. Periodic boundary conditions are imposed on the endpoints of the domain. The resulting system is integrated using the MATLAB ODE solver ode15s. In (a), $A=1.5$ and $\chi=0.2$ and the system is initially perturbed randomly from a state in which all densities are uniform in space. In (b) $A=2.4$ and $\chi=0.8$ and the simulation is started from a tree-pattern to which a random perturbation is added. Both initial conditions are obtained from results of the one-species extended Klausmeier model \eqref{eq: Multispecies: Model: Klausmeier}. The other parameter values in all of the figures are $B_1=0.45$, $b_2 = 0.0055$, $f=0.01$, $h=0.01$, $s=10^{-3}$, $d=500$ and $M=2^{9}$.}\label{fig: Multispecies: Simulations: example solutions coex patterns}
\end{figure}

In the parameter setting \eqref{eq: Multispecies: Model: parameter region to compare species} two distinct initial configurations from which such metastable states arise are established. If the initial condition is set to a state in which both plant species and the water density are uniform in space with a random perturbation added, then the solution remains in a metastable configuration in which both species coexist for a long time. If the rainfall is sufficiently low, the solution develops a patterned appearance in all three variables during the long transient. Eventually the metastable state reduces to a stable single-species equilibrium. The type of this equilibrium depends on the choice of parameters and, in particular, on the level of rainfall (see Figure \ref{fig: Multispecies: Simulations: example solutions coex patterns initally coex ss}). A sufficiently high level of rainfall leads to a spatially uniform solution, while lower amounts of precipitation cause convergence to a single-species pattern. The initial densities for the uniform state are chosen based on the steady states of the one-species Klausmeier model \eqref{eq: Multispecies: Model: Klausmeier} \cite{Klausmeier1999}.

A similar behaviour is exhibited by the model's solution if the initial condition of the system is set to a tree-only pattern that is obtained from the one-species Klausmeier model \eqref{eq: Multispecies: Model: Klausmeier}. To this configuration a low density of the grass variable $u_1$ is added, as well as a random perturbation in all three variables. In this scenario, the grass density $u_1$ quickly adopts a pattern that is in phase with the tree density $u_2$. The solution remains in this configuration for a long time, but a sharp reduction in tree density and changes to the wavelength of the pattern may occur. Eventually a transition to a grass-only equilibrium occurs. As described above, the choice of this grass-only equilibrium to which the system eventually converges depends on the precipitation parameter $A$ (see Figure \ref{fig: Multispecies: Simulations: example solutions coex patterns initially tree pattern}).

Such metastable patterns are not only observed for the parameter values chosen in Figure \ref{fig: Multispecies: Simulations: example solutions coex patterns}, but occur for a wide range of parameters. This motivates a closer investigation of the coexistence patterns and, in particular, their metastability. One possibility to gain a comprehensive understanding of the patterns' properties would be a systematic numerical investigation of the whole parameter space. Such an approach could involve the tracking of the time the system spends in the coexistence state under variations of both single parameters and combinations of multiple parameters, as well as a closer investigation of the pattern's properties such as its wavelength . However, the number of different parameters in the model poses a significant challenge for this approach. Instead, linear stability analysis can be used to study the existence and stability of such patterns, which is presented in Sections \ref{sec: Multispecies: Metastable patterns from tree patterns} and \ref{sec: Multispecies: Metastable patterns from coexistence steady state}.

\section{Metastable coexistence patterns arising from stable one-species patterns}\label{sec: Multispecies: Metastable patterns from tree patterns}

A common tool to study pattern formation in reaction-diffusion systems is linear stability analysis. Motivated by the simulation visualised in Figure \ref{fig: Multispecies: Simulations: example solutions coex patterns initially tree pattern}, we use linear stability analysis to discuss the emergence of metastable patterns in which both species coexist from a stable one-species Turing-type pattern into which a new species is introduced.

Linear stability analysis is based on the growth/decay of perturbations to equilibria of the system. Depending on the choice of parameters \eqref{eq: Multispecies: Model: species same model} has up to seven spatially homogeneous steady states; a trivial state describing desert which is stable in the whole parameter space, and pairs of semi-trivial single-species steady states as well as a pair of equilibria that correspond to coexistence of both species. To differentiate between the two types of patterns addressed in this section, we strictly refer to a pattern to be of Turing-type if it emerges from a steady state that is linearly stable to spatially uniform perturbations and becomes unstable upon introduction of spatial variation in the perturbation. An equilibrium of \eqref{eq: Multispecies: Model: species same model} is linearly stable to spatially homogeneous perturbations if all eigenvalues $\lambda_u \in \C$ of the system's Jacobian at the steady states satisfy $\Re(\lambda_u)<0$. For \eqref{eq: Multispecies: Model: nondimensional model}, the Jacobian is given by $J(u_1,u_2,w) = (j(u_1,u_2,w)_{k\ell})$, $k,\ell =1,2,3$, where
\begin{align}
\begin{split}\label{eq: Multispecies: LinStab: Jacobian}
j\left(u_1,u_2,w\right)_{11}  &= \left(Hw-S\right)u_2+2u_1w-B_1, \\ 
j\left(u_1,u_2,w\right)_{12}  &= u_1\left(Hw-S \right), \\ 
j\left(u_1,u_2,w\right)_{13}  &=  u_1\left(u_1+Hu_2\right), \\ 
j\left(u_1,u_2,w\right)_{21}  &= Fu_2w, \\ 
j\left(u_1,u_2,w\right)_{22}  &= 2Fw\left(\dfrac{u_1}{2}+Hu_2 \right)-B_2, \\ 
j\left(u_1,u_2,w\right)_{23}  &= Fu_2\left(u_1+Hu_2 \right), \\ 
j\left(u_1,u_2,w\right)_{31}  &= -w\left(2u_1+\left(1+H\right)u_2 \right), \\ 
j\left(u_1,u_2,w\right)_{32}  &= -w\left(\left(1+H \right)u_1+2Hu_2 \right), \\ 
j\left(u_1,u_2,w\right)_{33} &= -u_1^2-\left(1+H\right)u_1u_2-Hu_2^2-1.
\end{split}
\end{align}
For an equilibrium that is linearly stable to spatially uniform perturbations, Turing-type patterns emerge if there exists a wavenumber $k>0$ such that at least one eigenvalue $\lambda_s \in \C$ of $J-\operatorname{diag}(k^2, Dk^2, dk^2)$ has positive real part, i.e. $\max_{k\ge0,\lambda_s}\{\Re(\lambda_s)\}>0$.

Although $\max_{k\ge0,\lambda_s}\{\Re(\lambda_s)\}>0$ is a necessary condition for the development of a pattern from a spatial perturbation, $\max_{\lambda_u}\{\Re(\lambda_u)\}<0$ is not necessarily required. Spatial patterns also form if  $0<\max_{\lambda_u}\{\Re(\lambda_u)\}\ll \max_{k\ge0,\lambda_s}\{\Re(\lambda_s)\}$. In this case a pattern (and the corresponding equilibrium) is unstable but the difference in the growth rates gives rise to a transient pattern but the solution eventually tends to a stable state. In particular, if $\max_{\lambda_u}\{\Re(\lambda_u)\}\ll 1$, this transient occurs at a slow rate as visualised in Figure \ref{fig: Multispecies: Simulations: example solutions coex patterns} and the pattern is metastable.

\subsection{Turing-type patterns}

Investigation of the existence of such metastable patterns requires a understanding of the model's single-species Turing-type patterns. Due to the nature of the model, the linear stability analysis of the single-species equilibria is almost identical to that of the extended Klausmeier model on flat ground, in which patterns emerge from a Turing bifurcation. The considerations for \eqref{eq: Multispecies: Model: nondimensional model} only differ from those of the Klausmeier model through the existence of an additional condition that determines the stability to the introduction of the second species. Moreover, in case of the tree-only equilibria the parameters $F$, $H$ and $D$ alter the stability conditions quantitatively.

For each plant species, there exists a pair of semi-trivial steady states in which only one plant species prevails. Provided $A>A_{\min}^G:=2B_1$, the grass equilibrium is 
\begin{align*}
\left(\overline{u}_1^{G,\pm}, 0, \overline{w}^{G,\pm}\right) = \left(\frac{A\pm\sqrt{A^2-4B_1^2}}{2B_1},0,\frac{2B_1^2}{A\pm\sqrt{A^2-4B_1^2}}\right),
\end{align*}
where the superscript $G$ identifies it as a single-species grass state and $\pm$ indicates the choice of sign. Similarly, the pair of steady states describing a tree-only state is given by 
\begin{align*}
\left(0,\overline{u}_2^{T,\pm},  \overline{w}^{T,\pm}\right) =  \left(0,\frac{\xi_\pm}{2B_2H},\frac{2B_2^2}{F\xi_\pm}\right),
\end{align*}
provided the precipitation parameter exceeds $A_{\min,\operatorname{ex}}^T:=2B_2F^{-1}H^{-(1/2)}$, where $\xi_\pm = AFH\pm\sqrt{A^2F^2H^2-4B_2^2H}$.

\subsubsection{Stability to Spatially Uniform Perturbations}
The initial step in determining conditions for the existence of Turing-type patterns is linear stability analysis in a spatially uniform setting. Assuming no space dependence in \eqref{eq: Multispecies: Model: nondimensional model}, an equilibrium's stability is determined by the eigenvalues of the Jacobian with entries \eqref{eq: Multispecies: LinStab: Jacobian} evaluated at the equilibrium. For the grass-only steady state $(\overline{u}_1^{G,\pm}, 0, \overline{w}^{G,\pm})$ the Jacobian is
\begin{align*}
J^{G,\pm} = \matthree{B_1}{\dfrac{2B_1^2H-SA-S\sqrt{A^2-4B_1^2}}{2B_1}}{\dfrac{\left(A\pm\sqrt{A^2-4B_1^2}\right)^2}{4B_1^2}}{0}{B_1F-B_2}{0}{-2B_1}{-B_1\left(1+H\right)}{-\dfrac{A\left(A\pm\sqrt{A^2-4B_1^2}\right)}{2B_1^2}}.
\end{align*} 
Thus, the eigenvalues $\lambda_u^{G,\pm} \in \C$ satisfy
\begin{align}\label{eq: Multispecies: LinStab eigenvalues grass only steady state}
\left(B_1F-B_2 -\lambda_u^{G,\pm} \right) \det\mattwo{B_1-\lambda_u^{G,\pm}}{\dfrac{\left(A\pm\sqrt{A^2-4B_1^2}\right)^2}{4B_1^2}}{-2B_1}{-\dfrac{A\left(A\pm\sqrt{A^2-4B_1^2}\right)}{2B_1^2}-\lambda_u^{G,\pm}} = 0.
\end{align}
The eigenvalue $\lambda_{u,1}^{G,\pm} := B_1F-B_2$ accounts for the introduction of the tree species $u_2$, while the remaining two eigenvalues are independent of any parameters associated with $u_2$. Indeed, the matrix in \eqref{eq: Multispecies: LinStab eigenvalues grass only steady state} is identical to that of the corresponding matrix obtained in the linear stability analysis of the Klausmeier model in which only a single species is considered. Thus $(\overline{u}_1^{G,+}, 0, \overline{w}^{G,+})$ is linearly stable to spatially homogeneous perturbations if $A>A_{\min}^G$, $B_2>B_1F$ and $B_1<2$, while $(\overline{u}_1^{G,-}, 0, \overline{w}^{G,-})$ is linearly unstable for any choice of parameters \cite{Klausmeier1999,Sherratt2005}.

Similar to the analysis of the grass steady state, the tree equilibrium $(0,\overline{u}_2^{T,-}, \overline{w}^{T,-})$ is linearly unstable in the whole parameter space and $(0, \overline{u}_2^{T,+},  \overline{w}^{T,+})$ is linearly stable to spatially homogeneous perturbations if $A>A_{\min,\operatorname{ex}}^T$, $B_2<2$ and 
\begin{align}\label{eq: Multispecies: LinStab: shading threshold for stability}
S>\frac{2B_2H\left(B_2-B_1F \right)}{F\xi_+}.
\end{align}
Similar to the stability conditions of the single-species grass equilibrium, only criterion \eqref{eq: Multispecies: LinStab: shading threshold for stability} accounts for the stability of $(0, \overline{u}_2^{T,+},  \overline{w}^{T,+})$ to the introduction of $u_1$. Thus, the stable (provided $A>A_{\min,\operatorname{ex}}^T$ and $B_2<2$) single-species tree equilibrium becomes unstable to perturbations in the grass variable $u_1$ if the shading parameter is sufficiently small (see the difference between Figures \ref{fig: Multispecies: LinStab: stability diagram change all parameters S 1} and \ref{fig: Multispecies: LinStab: stability diagram change all parameters S small}).  Rearranging \eqref{eq: Multispecies: LinStab: shading threshold for stability} and combining it with the threshold $A_{\min,\operatorname{ex}}^T$ for existence of the steady state yields that $(0, \overline{u}_2^{T,+},  \overline{w}^{T,+})$ exists and is linearly stable if $B_2<2$ and 
\begin{align}\label{eq: Multispecies: LinStab: Amin tree steady state}
A>A_{\min}^T := \begin{cases}
\dfrac{2B_2}{F\sqrt{H}} \quad &\text{if} \quad S>S_c\\
\dfrac{B_2\left(\left(B_1^2H+S^2 \right)F^2-2B_1B_2FH+B_2^2H \right)}{\left(B_2-B_1F \right)F^2HS} \quad &\text{if} \quad S<S_c \\
\end{cases},
\end{align}
where $S_c:=\sqrt{H}(B_2-B_1F)F^{-1}$. This lower bound is derived through calculation of the eigenvalues $\lambda_{u}^{T,\pm} \in \C$ of the Jacobian at $(0, \overline{u}_2^{T,\pm}, \overline{w}^{T,\pm})$ which satisfy
\begin{align*}
\left(\frac{2B_2H\left(B_2-B_1F \right)-SF\xi_\pm}{2FHB_2} -\lambda_{u}^{T,\pm}\right)  \det\left(J^{T,\pm}-\lambda_{u}^{T,\pm} I_2 \right) =0,
\end{align*}
where
\begin{align*}
J^{T,\pm} =  \mattwo{B_2}{\dfrac{F\xi_\pm^2}{4B_2^2H}}{-\dfrac{2B_2}{F}}{-\dfrac{AF\xi_\pm}{2B_2^2} },
\end{align*}
and $I$ is the identity matrix. Imposing a negativity condition on the root $\lambda_{u}^{C,+}$ given by the first factor of this product yields \eqref{eq: Multispecies: LinStab: shading threshold for stability}, while the remaining two eigenvalues are both negative if and only if $\tr(J^{T,\pm})<0$ and $\det(J^{T,\pm})>0$. For $(0, \overline{u}_2^{T,-}, \overline{w}^{T,-})$, $\det(J^{T,-})<0$ for any choice of parameters yielding its instability, while for $(0, \overline{u}_2^{T,+}, \overline{w}^{T,+})$, $\det(J^{T,+})>0$. Finally, stability requires $\tr(J^{T,+})>0$ which holds for all $B_2<2$.

Bistability of the tree-only steady state and the grass-only steady state requires stability of both semi-trivial equilibria to the introduction of the other species. Stability of the single-species grass equilibrium $(\overline{u}_1^{G,+}, 0, \overline{w}^{G,+})$ to the introduction of the tree species $u_2$, i.e. $B_2>B_1F$, occurs if the grass species has a superior water to biomass conversion to mortality rate, which we define to be a measure of a species' average fitness. To balance this disadvantage, stability of the tree-only state $(0,\overline{u}_2^{T,+}, \overline{w}^{T,+})$ to the introduction of the grass species $u_1$ necessitates the shading effect to be sufficiently large. Indeed, if $B_2>B_1F$ and $S<S_c$, then 
\begin{align*}
A_{\min}^T =\frac{B_2\left(\left(B_1^2H+S^2 \right)F^2-2B_1B_2FH+B_2^2H \right)}{\left(B_2-B_1F \right)F^2HS},
\end{align*}
which is decreasing in $S$ below the threshold $S_c$. Thus, in the parameter region in which the grass-only steady state is stable, a decrease in the inhibitory shading effect of trees on grass increases the precipitation requirement for bistability of the tree-only and grass-only steady state. This is visualised in Figures \ref{fig: Multispecies: LinStab: stability diagram change all parameters S 1} and \ref{fig: Multispecies: LinStab: stability diagram change all parameters S small}. The threshold $S_c$ defined in \eqref{eq: Multispecies: LinStab: Amin tree steady state}, which is of the same order of magnitude as the average fitness difference $B_2-B_1F$ between the species, describes the intensity of shading above which the tree equilibrium $(0,\overline{u}_2^{T,+}, \overline{w}^{T,+})$ is stable to the introduction of the grass-species $u_1$ for any precipitation levels that guarantee the existence of the steady state. In other words, if the shading effect of $u_2$ on $u_1$ is sufficiently large, then $(0,\overline{u}_2^{T,+}, \overline{w}^{T,+})$ is always linearly stable to the introduction of the grass-species $u_1$.

The bounds on the plant mortality parameters in the derivations above are sufficient but not necessary. However, parameter estimates consistently indicate that $B_1<2$ and $B_2<2$ \cite{Klausmeier1999, Synodinos2015} and thus we restrict the analysis to this region.

\subsubsection{Conditions for the formation of Turing-type patterns}\label{sec: Multispecies: LinStab: conditions for pattern formation}
Having established stability conditions for the single-species equilibria in a spatially uniform setting, we turn to spatially non-uniform perturbations of the steady states to determine the loci of Turing bifurcations. Typically, linear stability analysis is used to study pattern formation by introducing perturbations of the steady state that are proportional to $\exp(\lambda_s t+ikx)$ for a growth rate $\lambda_s\in\C$ and wavenumber $k>0$. Imposing such perturbations on the semi-trivial steady states, i.e. $(\overline{u}_1^{G,+}, 0, \overline{w}^{G,+})$ and $(0,\overline{u}_2^{T,+}, \overline{w}^{T,+})$, however, would yield negative plant densities, a biologically unrealistic scenario. To avoid this, the density of the species that vanishes at the steady state is kept at zero. This reduces the model to the one-species Klausmeier model with water diffusion (up to the constants $F$, $H$ and $D$ in case of $(0,\overline{u}_2^{T,+}, \overline{w}^{T,+})$), for which patterns form due to a diffusion-driven instability.

More precisely, for $(\overline{u}_1^{G,+}, 0, \overline{w}^{G,+})$ \eqref{eq: Multispecies: Model: nondimensional model} reduces to 
\begin{align*}
\frac{\partial u_1}{\partial t} &= wu_1^2-B_1 u_1 + \frac{\partial^2 u_1}{\partial x^2}, \\
\frac{\partial w}{\partial t} &= A-w - wu_1^2 +d \frac{\partial^2 w}{\partial x^2},
\end{align*}
which is the extended Klausmeier model on flat ground. The typical linear stability analysis approach outlined above yields that a pattern-forming instability occurs for
\begin{multline}\label{eq: Multispecies: LinStab: grass pattern condition}
A_{\min}^G<A<A_{\max}^{G,+} := \\ \frac{B_1^{\frac{3}{2}}d^{\frac{1}{2}}\left(3B_1^2d^2+7B_1d-8-2\sqrt{2B_1^4d^4+6B_1^3d^3-8B_1d} \right)^{\frac{1}{2}}}{dB_1+1},
\end{multline}
provided $d>B_1^{-1}$. If $d<B_1^{-1}$ then $A_{\max}^{G,+} \in \C$ and no Turing bifurcation occurs.

Similarly, setting $u_1=0$ in \eqref{eq: Multispecies: Model: nondimensional model}, i.e. considering the tree-only steady state $(0,\overline{u}_2^{T,+}, \overline{w}^{T,+})$, yields
\begin{subequations}\label{eq: Multispecies: LinStab: Tree only model}
	\begin{align}
	\frac{\partial u_2}{\partial t} &= FHwu_2^2-B_2 u_2 + D\frac{\partial^2 u_2}{\partial x^2}, \\
	\frac{\partial w}{\partial t} &= A-w - Hwu_2^2 +d \frac{\partial^2 w}{\partial x^2}.
	\end{align}
\end{subequations}
Considerations identical to those in the analysis of the extended Klausmeier model show that an instability leading to the formation of a tree pattern occurs if 
\begin{multline}\label{eq: Multispecies: LinStab: tree pattern condition}
A_{\min,\operatorname{ex}}^T<A<A_{\max}^{T,+} := \\ \frac{B_2^{\frac{3}{2}}d^{\frac{1}{2}}\left(3B_2^2d^2+7DB_2d-8D^2-2\sqrt{2B_2^4d^4+6DB_2^3d^3-8D^3B_2d} \right)^{\frac{1}{2}}}{D^{\frac{1}{2}}FH^{\frac{1}{2}}\left(dB_2+D\right)},
\end{multline}
provided $d>DB_2^{-1}$. If $d<DB_2^{-1}$, then $A_{\max}^{T,+}\in \C$ and no Turing bifurcation occurs. 

\begin{figure}
	\centering
	\subfloat[\label{fig: Multispecies: LinStab: stability diagram change all parameters S 1}]{\includegraphics[height=0.48\textwidth]{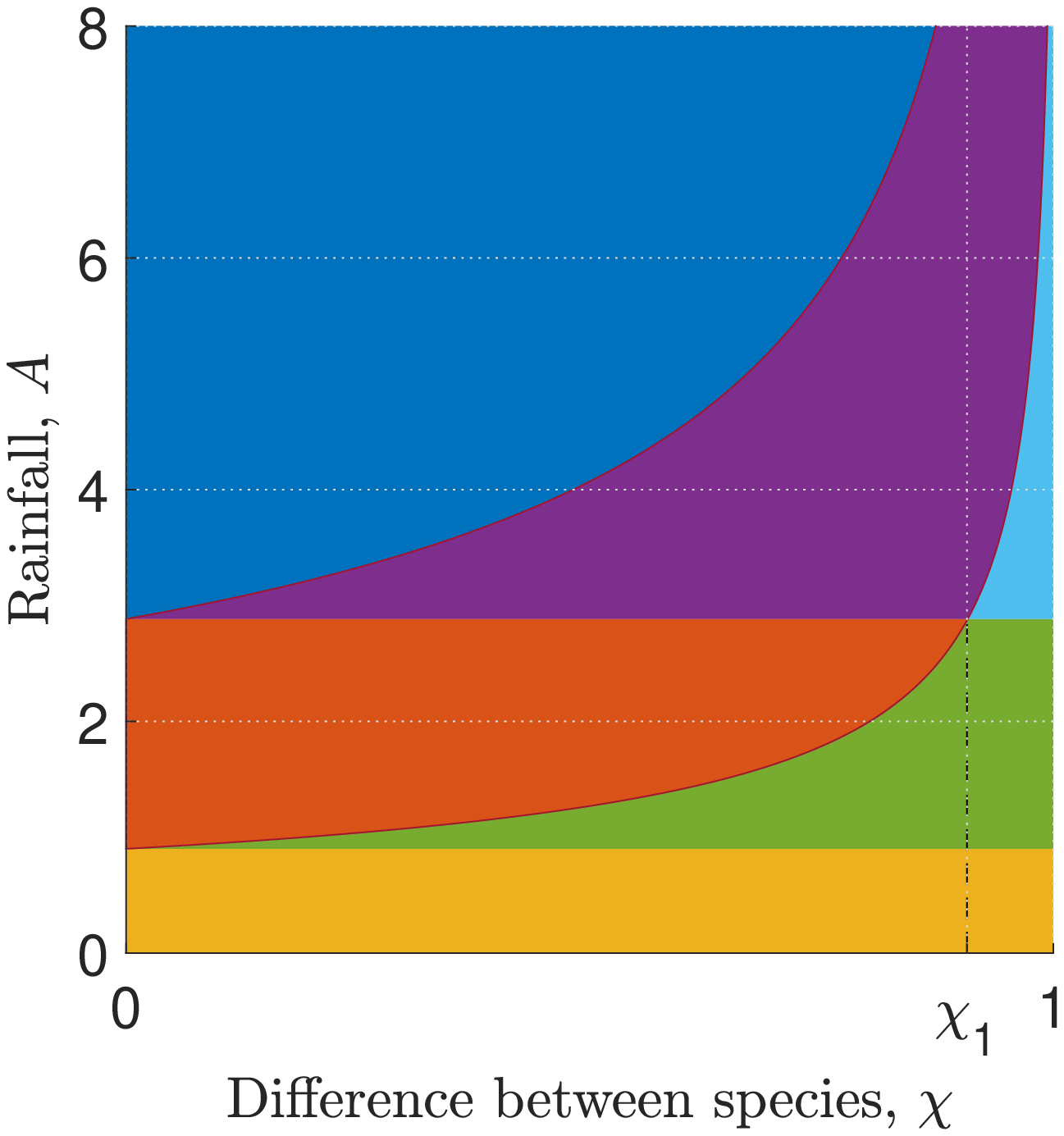}}
	\subfloat[\label{fig: Multispecies: LinStab: stability diagram change all parameters S 1 Dconst}]{\includegraphics[height=0.48\textwidth]{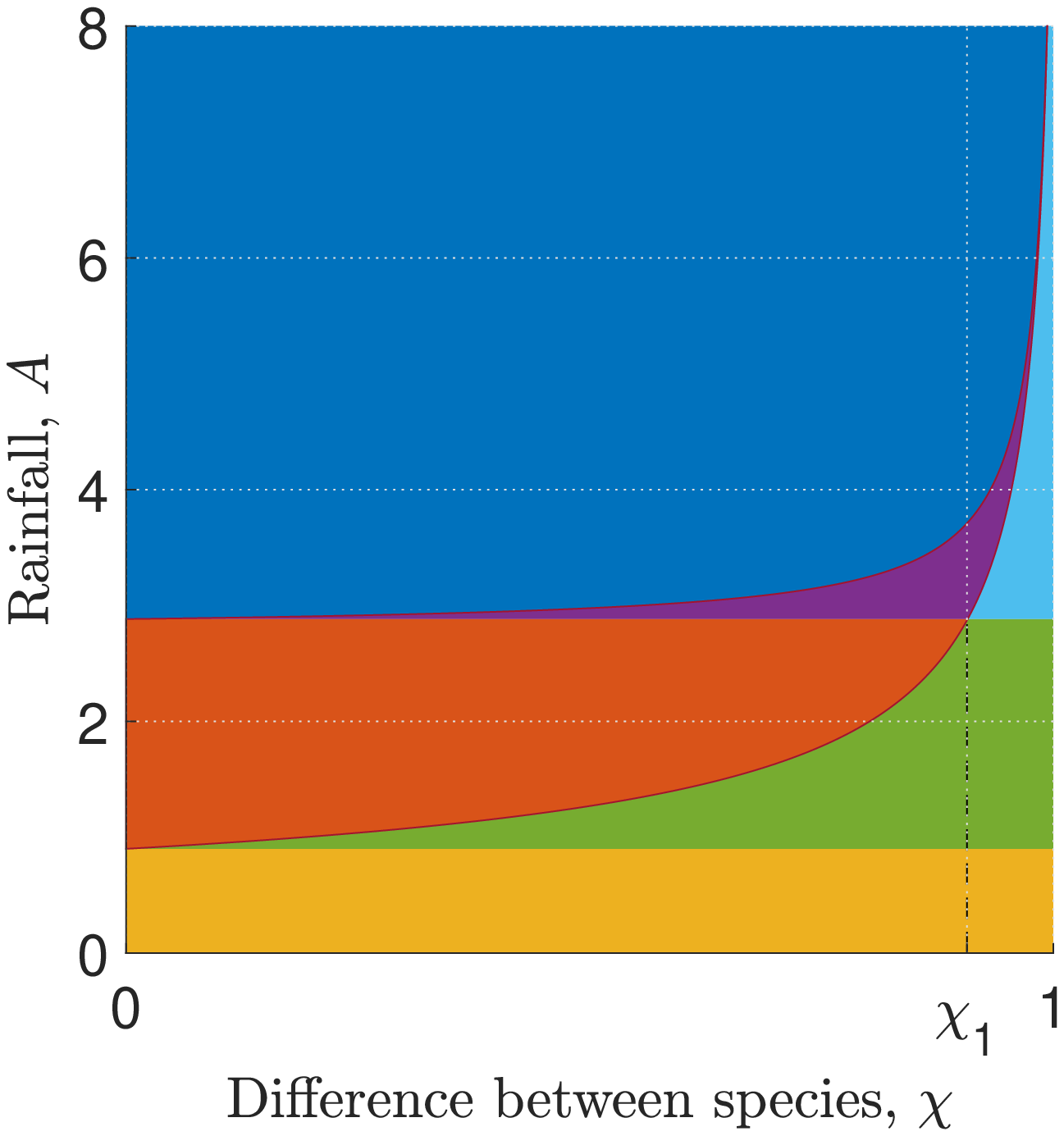}}\\
	\subfloat[\label{fig: Multispecies: LinStab: stability diagram change all parameters S small}]{\includegraphics[height=0.48\textwidth]{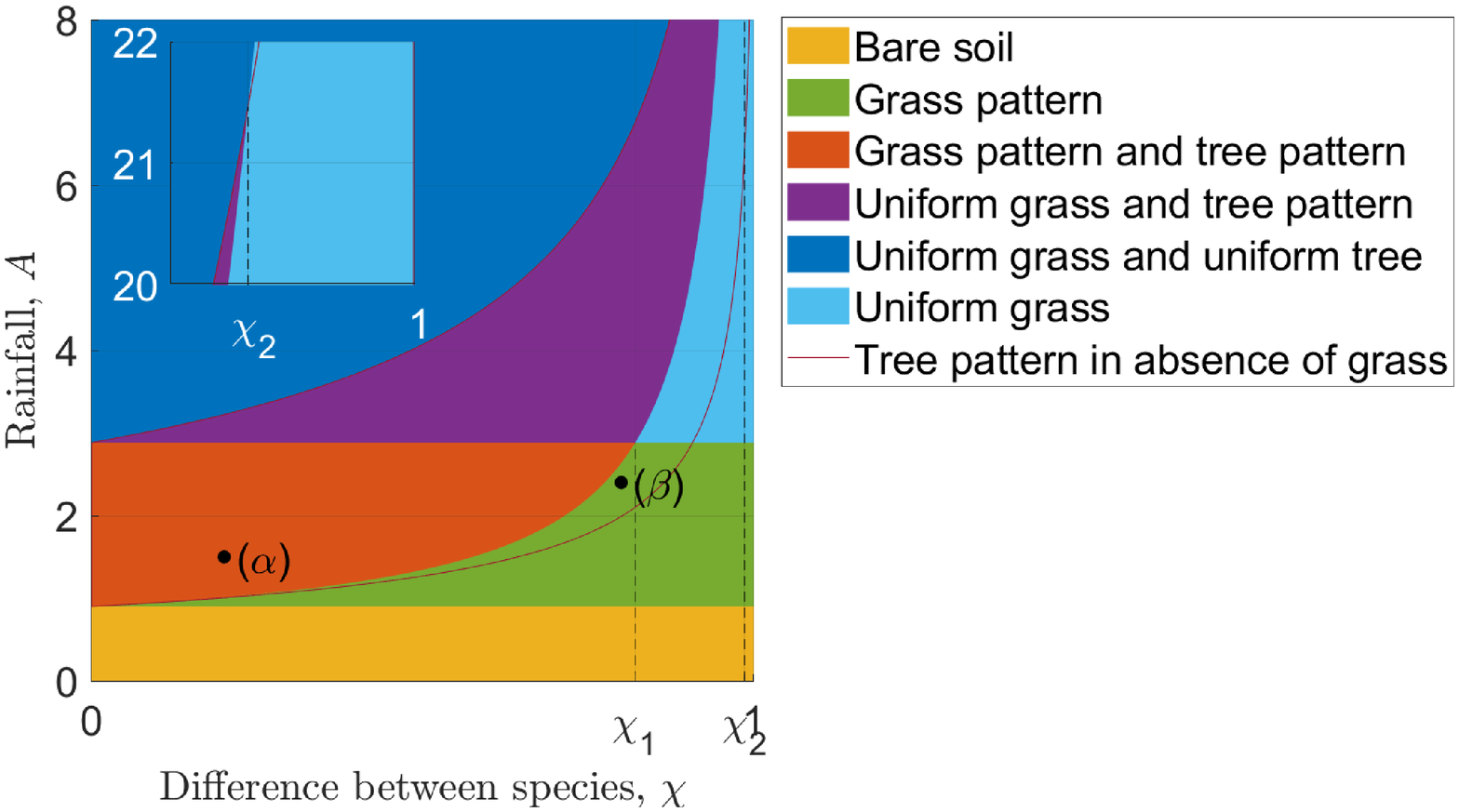}}
	\caption{Stability diagram for the semi-trivial steady states. The coloured areas combine the results of the linear stability analysis of the full model to spatially homogeneous perturbations and the respective one-species models in which spatially heterogeneous perturbations of the semi-trivial steady states lead to patterns in the parameter region \eqref{eq: Multispecies: Model: parameter region to compare species}. The solid line indicates the parameter region in which tree patterns form in the one-species model \eqref{eq: Multispecies: LinStab: Tree only model}. The difference between (a) and (c) shows that this does not coincide with the corresponding parameter region in the multispecies model if $S$ is small. The desert steady state is stable in the whole parameter plane. The area indicated in the figure only shows the region in which it is the only stable state. In (a) $s=1$ and $D=1-\chi(1-0.01)$, which gives $S<S_c$ for all $0<\chi<1$; in (b) $s=1$ and $D=1$; and in (c) $s=10^{-3}$ and $D=1-\chi(1-0.01)$. The inset in (c) shows the behaviour around $\chi = \chi_2$. The other parameter values in all of the figures are $B_1=0.45$, $b_2 = 0.0055$, $f=0.01$, $h=0.01$ and $d=500$. The markers $(\alpha)$ and $(\beta)$ refer to the parameter values used in the simulations presented in Figure \ref{fig: Multispecies: Simulations: example solutions coex patterns}.}\label{fig: Multispecies: LinStab: stability diagram change all parameters}
\end{figure}

Condition \eqref{eq: Multispecies: LinStab: tree pattern condition} is equivalent to the ratio $d/D$ of the diffusion coefficients exceeding a critical threshold. Thus, a lower rate of diffusion of the woody species increases the size of the parameter region supporting pattern formation. This phenomenon is visualised in the stability diagrams \ref{fig: Multispecies: LinStab: stability diagram change all parameters S 1} and \ref{fig: Multispecies: LinStab: stability diagram change all parameters S 1 Dconst}. It is important to emphasise that the bifurcation point $A_{\max}^{T,+}$ is obtained by considering perturbations in $u_2$ and $w$ only. The calculation of $A_{\max}^{T,+}$ does not take into account a possible introduction of the grass species $u_1$. Indeed, as the difference between Figure \ref{fig: Multispecies: LinStab: stability diagram change all parameters S 1} and \ref{fig: Multispecies: LinStab: stability diagram change all parameters S small} visualises, if the shading parameter $S$ is sufficiently small, then there exists a parameter region in which a single-species tree pattern is stable only in the context of a single-species model. The instability to an introduction of the grass species $u_1$ occurs due to an increase of  $A_{\min}^T$, given by \eqref{eq: Multispecies: LinStab: Amin tree steady state}, for decreasing $S$. For sufficiently small $S$ this causes $A_{\min}^T >A_{\min,\operatorname{ex}}^{T}$ and thus a tree-only pattern cannot form for $A_{\min,\operatorname{ex}}^{T}<A<A_{\min}^T$ if the assumption of $u_1=0$ is relaxed. Similarly, the pattern forming condition \eqref{eq: Multispecies: LinStab: grass pattern condition} obtained for $(\overline{u}_1^{G,+}, 0, \overline{w}^{G,+})$ only applies if the steady state is stable to perturbations in $u_2$, i.e. if $B_2>B_1F$. In the stability diagrams in Figure \ref{fig: Multispecies: LinStab: stability diagram change all parameters} a state is only assumed to occur if the introduction of the second species does not cause destabilisation. Even though this restricts the bistability region of both single-species equilibria, the numerical simulations presented in Section \ref{sec: Multispecies: Simulations} suggest that this restriction does not apply to metastable patterns in which both species coexist. In particular, the simulation visualised in Figure \ref{fig: Multispecies: Simulations: example solutions coex patterns initially tree pattern}, which corresponds to the $(\beta)$ marker in Figure \ref{fig: Multispecies: LinStab: stability diagram change all parameters S small}, lies outside the bistability region. Indeed, the parameter region $A_{\min,\operatorname{ex}}^T<A<A_{\min}^T$, i.e. the region in which the tree-pattern is stable in the one-species model but unstable to the introduction to the grass species, gives rise to a metastable pattern such as that shown in Figure \ref{fig: Multispecies: Simulations: example solutions coex patterns initially tree pattern} and is closely examined in Section \ref{sec: Multispecies: Coexistence patterns through tree patterns: Metastability}.

To address the effects caused by the difference between two plant species, we put particular emphasis on the parameter region given by \eqref{eq: Multispecies: Model: parameter region to compare species}, where the difference is described by a single parameter $0<\chi<1$ for simplicity. To focus on the possible coexistence of both plant types, we further restrict the parameter region to that of the grass-only steady state's stability, i.e. $A>2B_1$ and $B_2>FB_1$. The latter condition holds for all $0<\chi<1$ if $b_2>fB_1$.
The lowest levels of precipitation beyond the threshold $A=2B_1$ that separates the parameter region in which only the trivial desert equilibrium is stable from bistability or tristability regions of plant states and the bare soil state, only support grass patterns. For a sufficiently small difference $\chi<\chi_1$ between the grass and tree species, an increase of rainfall along the precipitation gradient leads to a region in which the two patterned states are stable, before the uniform grass-only steady state gains stability and eventually also the uniform tree equilibrium becomes stable to form a parameter region in which there is bistability of both uniform steady states. If the difference between the species is larger than the threshold $\chi_1$, then no bistability of both patterned states is possible. Instead, the uniform grass steady state becomes stable at rainfall levels that are lower than those required for a tree pattern to form (Figures \ref{fig: Multispecies: LinStab: stability diagram change all parameters S 1}, \ref{fig: Multispecies: LinStab: stability diagram change all parameters S 1 Dconst} and \ref{fig: Multispecies: LinStab: stability diagram change all parameters S small}). Finally, if $\chi>\chi_2>\chi_1$, where the threshold $\chi_2$ may be larger than unity, the system does not support the formation of tree patterns and there is a direct transition from the parameter region that supports only the uniform grass equilibrium to the region in which bistability of both uniform steady states occurs (Figure \ref{fig: Multispecies: LinStab: stability diagram change all parameters S small}).

\subsection{Metastable Patterns}\label{sec: Multispecies: Coexistence patterns through tree patterns: Metastability}
The results of the preceding linear stability analysis not only show the existence of single-species Turing patterns, but in the parameter region $A_{\min,\operatorname{ex}}^{T}<A<A_{\min}^T$ also that of metastable patterns, such as the pattern visualised in Figure \ref{fig: Multispecies: Simulations: example solutions coex patterns initially tree pattern}, in which both species coexist.

Provided it exists ($A>A_{\min,\operatorname{ex}}^{T}$), the tree-only equilibrium $(0,\overline{u}_2^{T,+}, \overline{w}^{T,+})$ is stable to spatially uniform perturbations in the tree density $u_2$ and the water density $w$ for all biologically relevant parameter values and tree patterns emerge from the steady state due to a Turing-type instability for sufficiently low precipitation levels. An additional stability condition \eqref{eq: Multispecies: LinStab: shading threshold for stability} arises from the introduction of the grass species $u_1$. If $(0,\overline{u}_2^{T,+}, \overline{w}^{T,+})$ is unstable to the introduction of $u_1$ ($A<A_{\min}^T$), the eigenvalue $\lambda_{u,1}^{T,+}$ corresponding to spatially uniform perturbations is of small size and thus gives rise to a metastable solution as shown in Figure \ref{fig: Multispecies: Simulations: example solutions coex patterns initially tree pattern}. If in addition $\Re(\lambda_{s,1}^{G,+}(k)) \gg \lambda_{u,1}^{T,+}$, where $\lambda_{s,1}^{T,+}(k) \in \C$ is the growth rate corresponding to a spatial perturbation with mode $k>0$, the grass species quickly (compared to the time it takes to reach the stable grass-only state) adopts a patterned appearance in phase with the tree pattern during this transition.  Indeed,
\begin{align}\label{eq: Multispecies: LinStab: unstable eigenvalue tree to grass introduction}
\lambda_{u,1}^{T,+} = \frac{2B_2H\left(B_2-B_1F\right)}{F\left(AFH + \sqrt{A^2F^2H^2-4B_2^2F^2H} \right)}-S \le \frac{2B_2}{AF^2} \left(B_2-B_1F\right)\ll 1,
\end{align} 
because tree mortality $B_2$ is of small size (see Table \ref{tab: Multispecies: Models: Parameters}). Further, the condition $\Re(\lambda_{s,1}^{G,+}(k)) \gg \lambda_{u,1}^{T,+}$ is satisfied unless parameter values are close to the grass-only steady state's Turing bifurcation locus. Thus, if a grass population is introduced into a stable tree pattern and causes destabilisation of this pattern as shown in Figure \ref{fig: Multispecies: Simulations: example solutions coex patterns initially tree pattern}, the small size of the eigenvalue (if positive) yields a slow transition to the stable grass-only state. 
The difference $B_2-B_1F$ plays a crucial role in the metastability property as it is the cause of the pattern's slow rate of destabilisation. Ecologically the small size of this difference corresponds to similar average fitness of both species. It is this balance that enables the coexistence of both species. The significance of $B_2-B_1F$ is not a special feature of this particular case but also causes the metastability of patterns originating from spatially uniform initial conditions such as that used in the simulation visualised in Figure \ref{fig: Multispecies: Simulations: example solutions coex patterns initally coex ss}. This is discussed in more detail in Section \ref{sec: Multispecies: Metastable patterns from coexistence steady state}.

Similar considerations suggest the possibility of metastable coexistence patterns that arise from the introduction of the tree species into a stable grass pattern that consequently becomes unstable. In this situation, however, the eigenvalue $\lambda_{u,1}^{G,+} = B_1F-B_2$ that corresponds to the introduction of the tree species is not necessarily small. Unless $\lambda_{u,1}^{G,+} \ll 1$, a perturbation of a grass pattern through the introduction of trees yields a quick transition to a tree pattern as a positive but not small value of $B_1F-B_2$ corresponds to a larger average fitness of the tree-species.

\subsubsection{Wavelength} \label{sec: Multispecies: LinStab: wavelength single}
A key feature of any regular pattern is its wavelength. While an extensive study of pattern wavelength requires tools from nonlinear analysis, linear stability analysis provides an insight into the wavelength of the patterns close to the bifurcation locus. Then the pattern wavelength is typically determined by the wavenumber that corresponds to the largest growth rate. Given such a wavenumber $k_{\max}$ calculated in the derivation of the Turing bifurcation points, the corresponding pattern has wavelength $L=2\pi/k_{\max}$.

\begin{figure}
	\centering
	\subfloat[Minimum wavelength.\label{fig: Multispecies: LinStab: single species pattern wavelengths min}]{\includegraphics[height=0.48\textwidth]{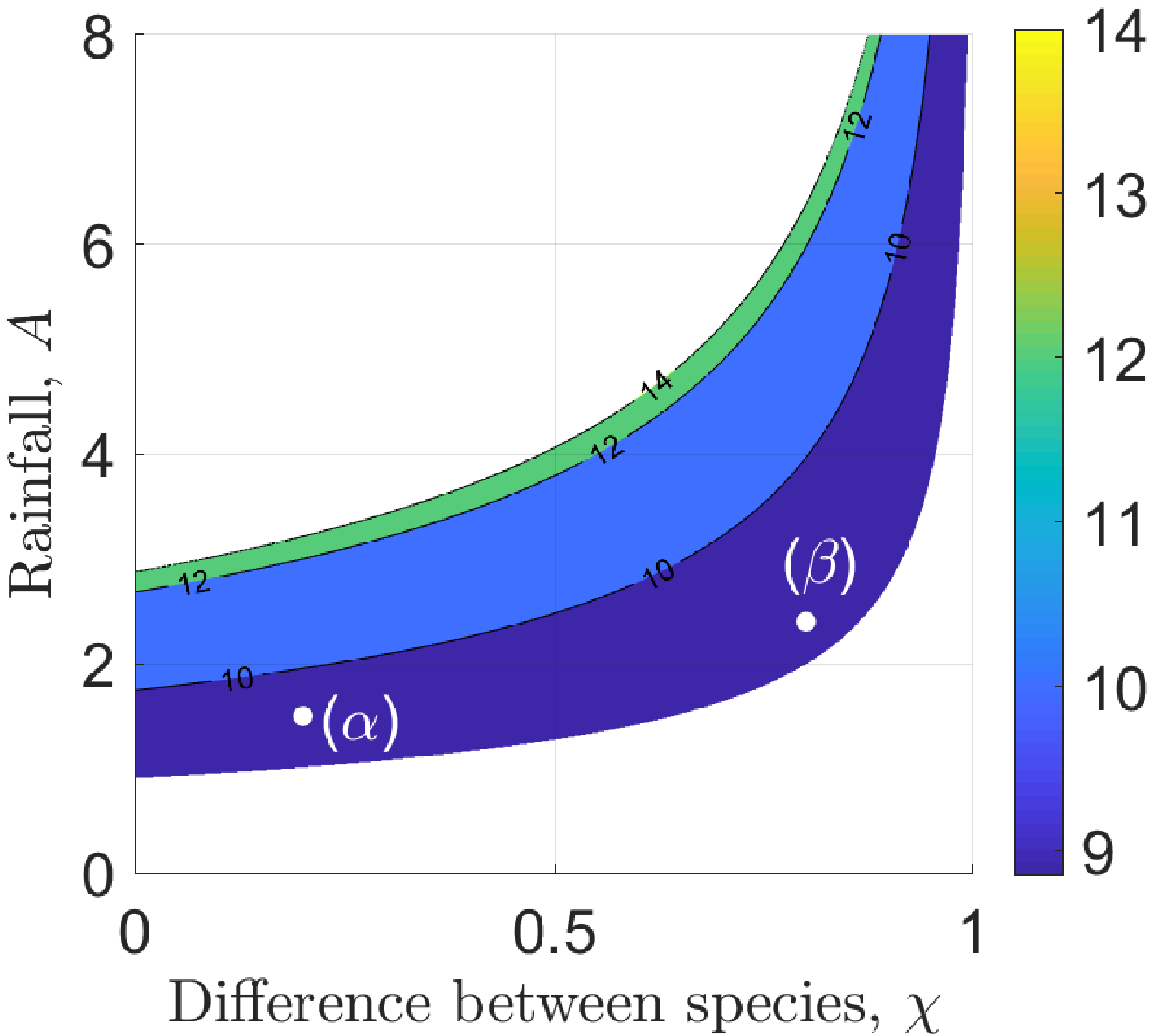}}
	\subfloat[Maximum wavelength.\label{fig: Multispecies: LinStab: single species pattern wavelengths max}]{\includegraphics[height=0.48\textwidth]{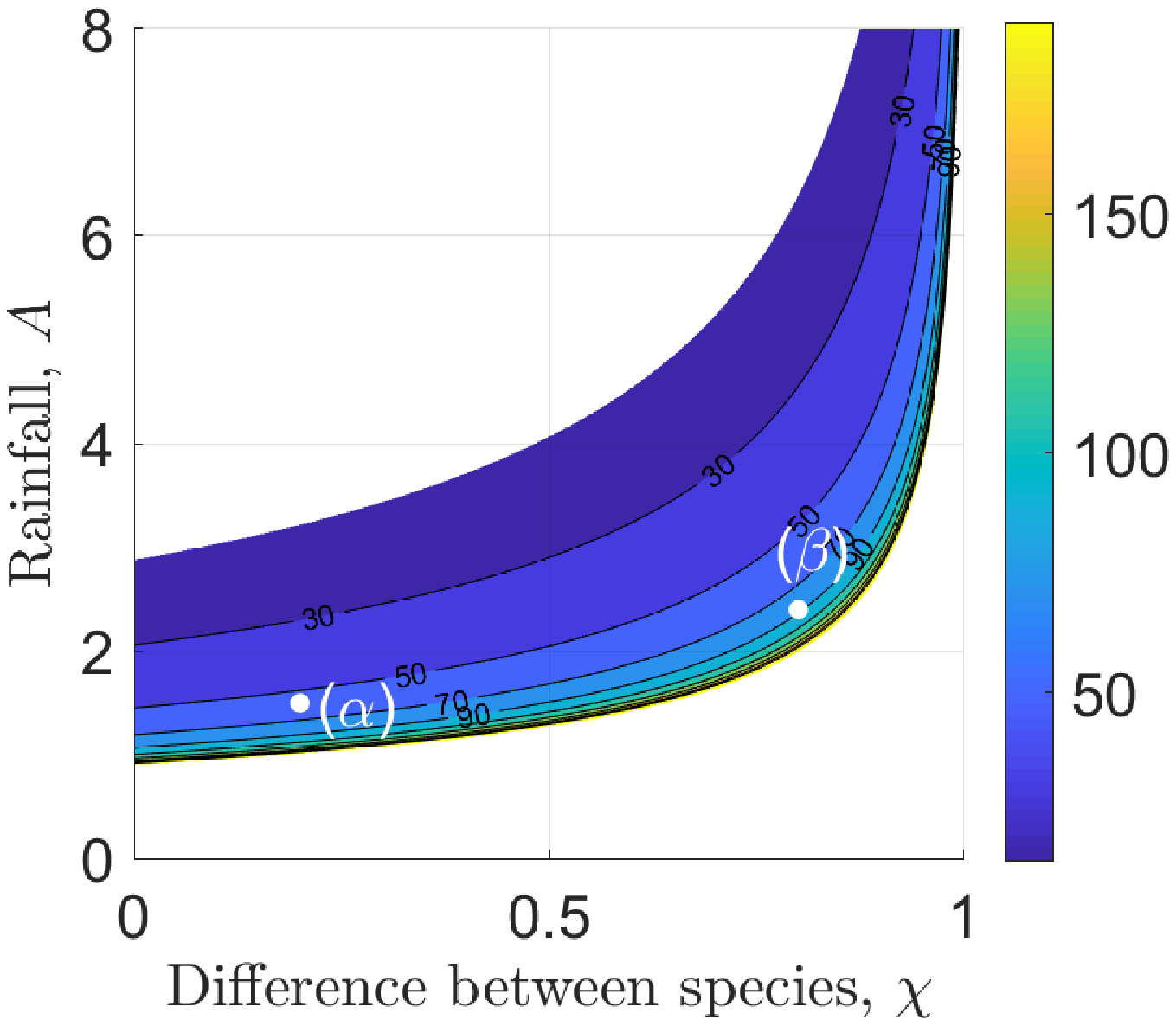}} \\
	\subfloat[Wavelength arising from most unstable wavenumber.\label{fig: Multispecies: LinStab: single species pattern wavelengths most unstable mode}]{\includegraphics[height=0.48\textwidth]{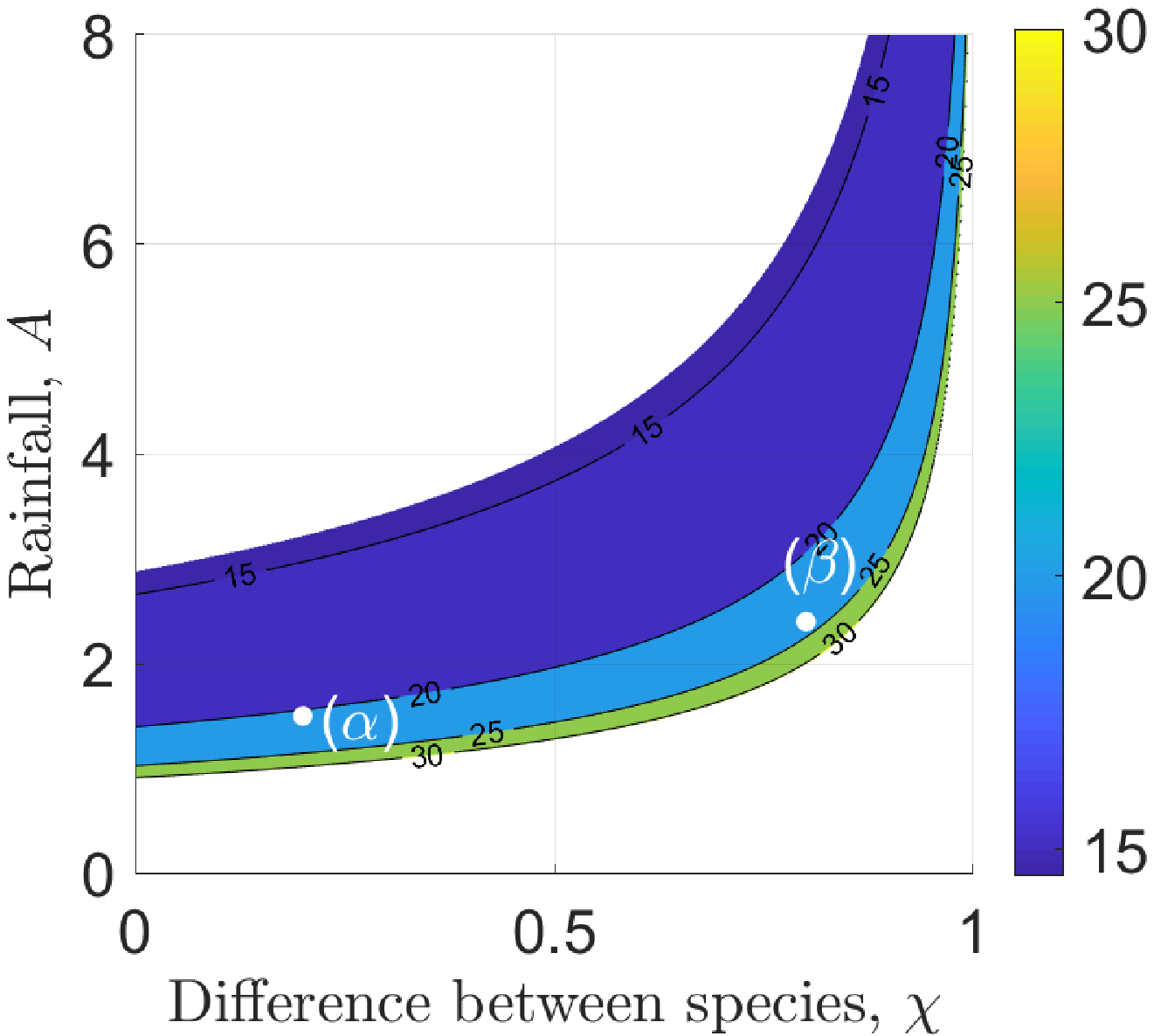}}
	\caption{Single species pattern wavelength. This figure visualises the pattern wavelengths of both single-species patterns calculated through linear stability analysis. The contours show the wavelength of the pattern of species $u_2$ as its difference from the grass species $u_1$ increases, while the values on the $A$-axis correspond to the wavelength of the grass pattern. Minimum (a), maximum (b) and wavelengths corresponding to the most unstable mode (c) are shown. The parameter values are $B_1=0.45$, $b_2 = 0.0055$, $f=0.01$, $h=0.01$, $D_0=0.01$, and $d=500$. The markers $(\alpha)$ and $(\beta)$ refer to the parameter values used in the simulations presented in Figure \ref{fig: Multispecies: Simulations: example solutions coex patterns}. For a comparison to the wavelengths of the coexistence pattern see Figure \ref{fig: Multispecies: LinStab: coex steady state wavelengths of patterns}.}\label{fig: Multispecies: LinStab: single species pattern wavelengths}
\end{figure}

From the preceding linear stability analysis we find that the wavelength of the tree species is increasing with the parameter $\chi$. Thus, for a constant level of precipitation, the more tree-like a species is, the longer is its pattern wavelength (Figure \ref{fig: Multispecies: LinStab: single species pattern wavelengths most unstable mode}). Such a comparison requires bistability of both patterned states, which is not necessarily the case for all $0\le \chi \le 1$, as indicated in  Figure \ref{fig: Multispecies: LinStab: single species pattern wavelengths}. The wavelength of both species further increases with decreasing rainfall, which is in agreement with results for the Klausmeier model on sloped terrain \cite{Sherratt2005, Sherratt2013IV}.

The most unstable wavenumber is not necessarily the mode that is selected in a pattern. Hysteresis is known to occur in the single-species Klausmeier model \cite{Sherratt2013, Siteur2014} and may cause the selected mode to differ from the most unstable mode. It is thus informative to obtain bounds on the wavelength from linear stability analysis. These bounds show that both an increase in species difference and lower precipitation increase the range of possible wavelengths (Figures \ref{fig: Multispecies: LinStab: single species pattern wavelengths min} and \ref{fig: Multispecies: LinStab: single species pattern wavelengths max}).

\section{Metastable coexistence patterns originate from a coexistence equilibrium} \label{sec: Multispecies: Metastable patterns from coexistence steady state}

The analysis in Section \ref{sec: Multispecies: Metastable patterns from tree patterns} only explains patterns in which both species coexist in the parameter region $A_{\min,\operatorname{ex}}^{T}<A<A_{\min}^T$. The simulations presented in Section \ref{sec: Multispecies: Simulations}, however, suggest that metastable coexistence patterns occur in a wider range of the precipitation parameter $A$. In this section we show that Turing-type patterns of the tree species $u_2$ are not the only origin of metastable patterns. Additionally, metastable patterns of species coexistence can arise from an equilibrium in which both species coexist, which is the subject of this section.

\begin{figure}
	\centering
	\subfloat[\label{fig: Multispecies: Equilibria: coex ss existence and positivity plus}]{\includegraphics[width=0.48\textwidth]{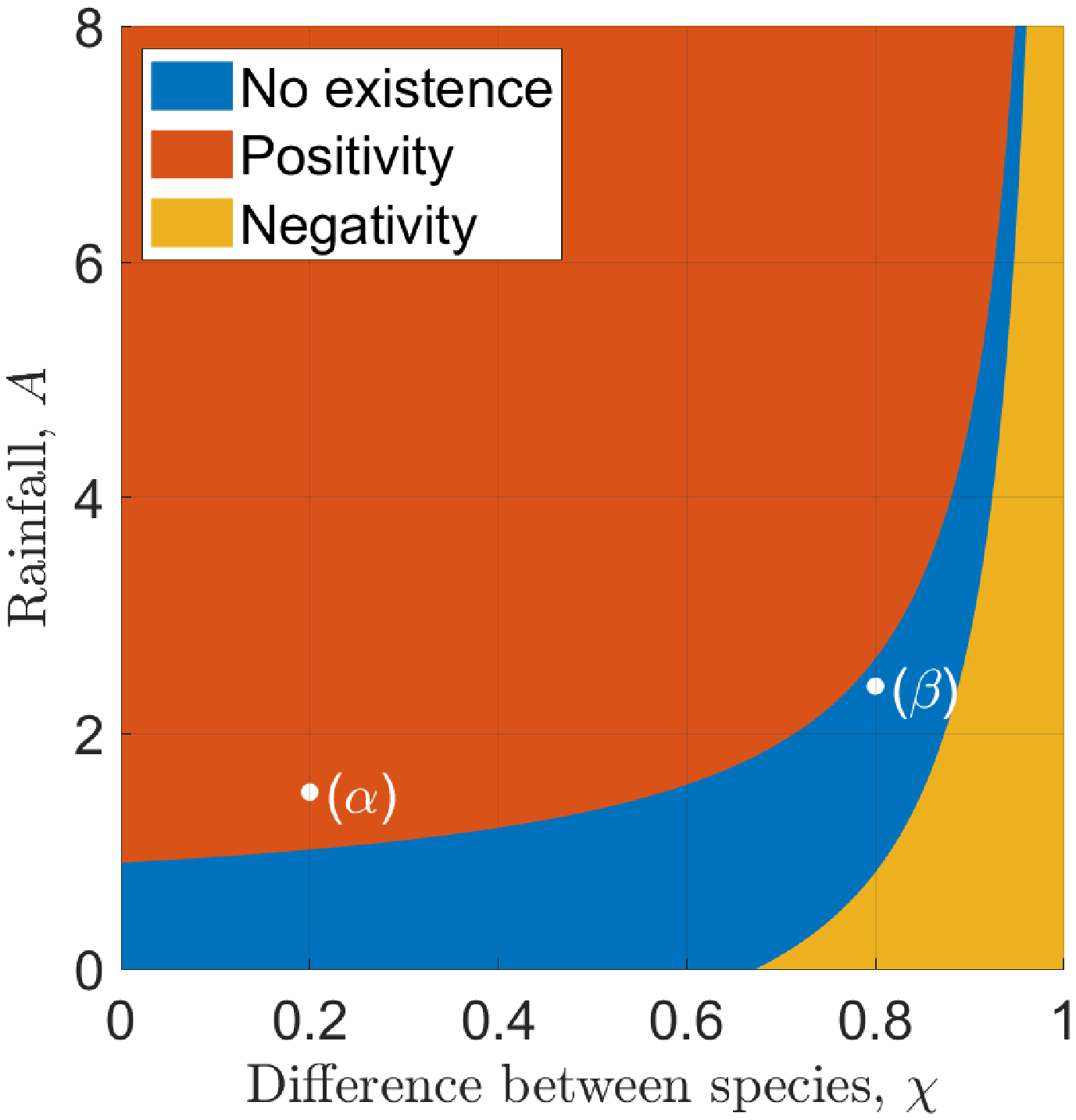}}
	\subfloat[]{\includegraphics[width=0.48\textwidth]{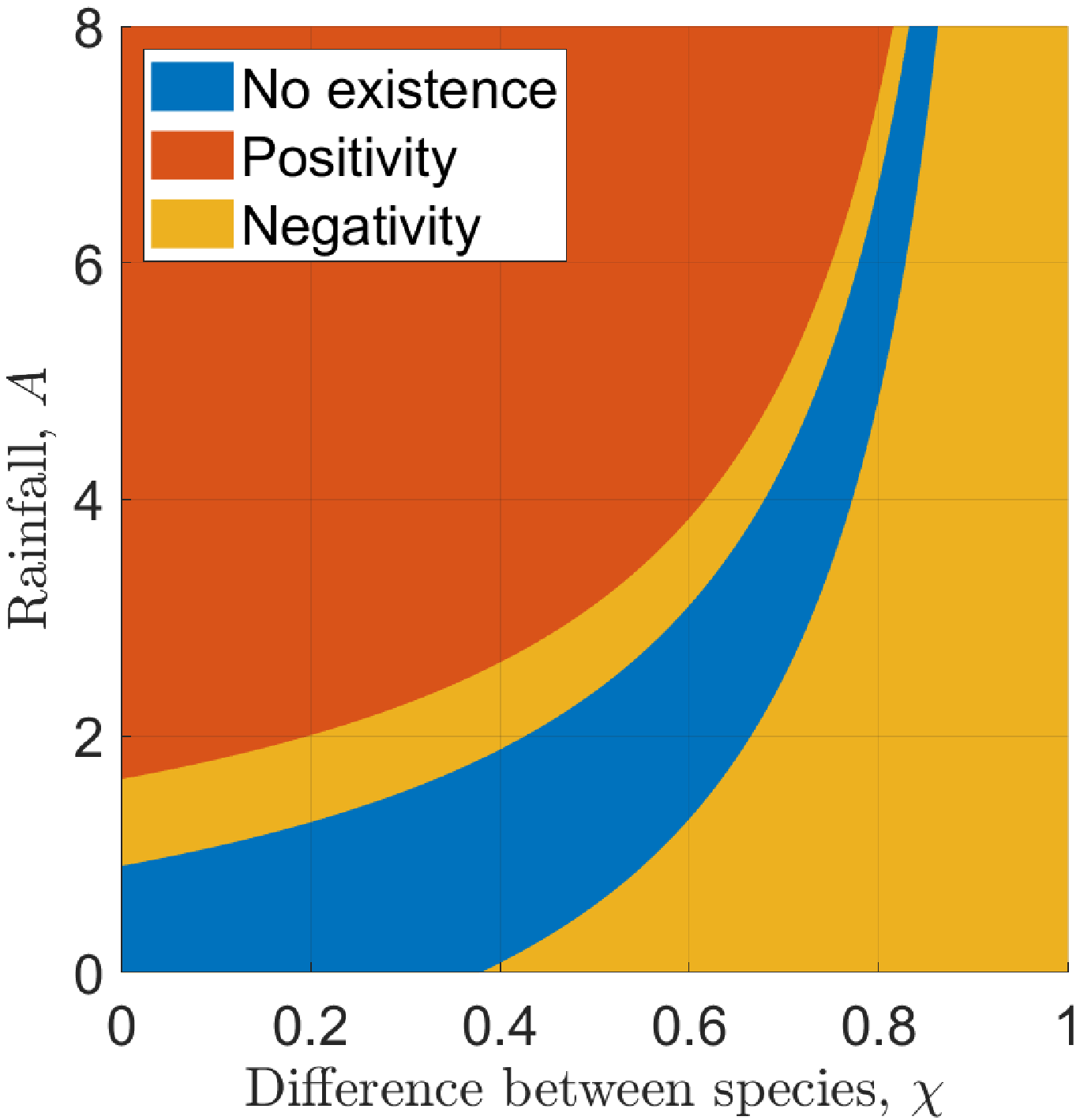}}
	\caption{Existence and positivity of the coexistence steady state. Visualisation of the parameter regions in which the coexistence steady state $(\overline{u}_1^{C,+}, \overline{u}_2^{C}, \overline{w}^{C,+})$ exists and is biologically relevant (positive) in the $\chi$-$A$ parameter plane for different levels of shading. In (a) $s=10^{-3}$, while in (b) $s=3\cdot 10^{-4}$. The other parameter values used in this visualisation are $B_1=0.45$, $b_2 = 0.0055$, $f=0.01$ and $h=0.01$. The legend of (a) also applies to (b). The markers $(\alpha)$ and $(\beta)$ in (a) refer to the parameter values used in the simulations presented in Figure \ref{fig: Multispecies: Simulations: example solutions coex patterns}.}\label{fig: Multispecies: Equilibria: coex ss existence and positivity}
\end{figure}

Besides the trivial and semi-trivial equilibria discussed in Section \ref{sec: Multispecies: Metastable patterns from tree patterns}, \eqref{eq: Multispecies: Model: nondimensional model} also admits a pair of coexistence steady states $(\overline{u}_1^{C,\pm}, \overline{u}_2^{C}, \overline{w}^{C,\pm})$, where similar to the notation used for the single species states the superscript $C$ identifies the equilibrium as a coexistence state. The equilibria satisfy
\begin{align*}
&\overline{u}_1^{C,\pm} = \frac{1}{2B_2} \left( AF-B_2\left(1+F\right)\overline{u}_2^{C} \right. \\ &\left. \pm \sqrt{\left(AF+B_2\left(1+F \right) \overline{u}_2^{C} \right)^2 - 4B_2\left(-AFH\overline{u}_2^{C} +B_2\left(1+H \left(\overline{u}_2^{C}\right)^2 \right) \right) } \right), \\
&\overline{u}_2^{C} = \frac{B_2-FB_1}{SF}, \quad \overline{w}^{C,\pm} = A-\frac{B_2}{F}\left(\overline{u}_1^{C,\pm}+\overline{u}_2^{C}\right),
\end{align*}
under suitable conditions that ensure their existence and biological relevance. For $(\overline{u}_1^{C,-}, \overline{u}_2^{C}, \overline{w}^{C,-})$ these are $B_2>B_1F$ and
\begin{align*}
\max \left\{\frac{B_2\left(\overline{u}_2^{C}\left(1-H\right)+2 \right)}{F}, \frac{B_2\overline{u}_2^{C}\left(1+F\right)}{F}\right\}<A<\frac{B_2\left(1+H\left(\overline{u}_2^{C}\right)^2 \right)}{FH\overline{u}_2^{C}} ,
\end{align*}
while the corresponding conditions for $(\overline{u}_1^{C,+}, \overline{u}_2^{C}, \overline{w}^{C,+})$ are $B_2>B_1F$ and
\begin{multline}\label{eq: Multispecies: LinStab: positivity and existence coex plus}
A>A_{\min}^{C,+}:= \\\max \left\{\frac{B_2\left(\overline{u}_2^{C}\left(1-H\right)+2 \right)}{F}, \min\left\{  \frac{B_2\overline{u}_2^{C}\left(1+F\right)}{F}, \frac{B_2\left(1+H\left(\overline{u}_2^{C}\right)^2 \right)}{FH\overline{u}_2^{C}}  \right\}\right\}.
\end{multline}
Visualisations in this paper are shown for the special parameter setting \eqref{eq: Multispecies: Model: parameter region to compare species} and $F=H$. In this situation changes to $\chi$ do not affect the nature of how the equilibrium loses its relevance. If $s>b_2-B_1f$, then $(\overline{u}_1^{C,+}, \overline{u}_2^{C}, \overline{w}^{C,+})$ ceases to exist at $A=A_{\min}^{C,+}$, while otherwise $A=A_{\min}^{C,+}$ represents the threshold at which  $\overline{u}_1^{C,+}$ becomes negative (see Figure \ref{fig: Multispecies: Equilibria: coex ss existence and positivity}). Similar considerations hold for $(\overline{u}_1^{C,-}, \overline{u}_2^{C}, \overline{w}^{C,-})$. This equilibrium, however, does not exhibit the metastability property which is the main focus of this paper and is therefore not considered further. It is noteworthy that there is nothing special about the choice of $F=H$ and results are robust to changes in $F$ and $H$, provided the rainfall minimum $A_{\min}^{C,+}$ remains in the biologically relevant parameter region. Results presented in this paper are also robust to changes in $s$.  Finally, we remark that the size of the shading parameter $S$ needs to be similar to that of the average fitness difference between both species $B_2-B_1F$ for the equilibrium to remain in a biologically relevant region, as large (small) shading effects only support coexistence at equilibrium if the density of $u_2$ is low (high).

An initial conclusion that is drawn from calculation of the existence region of the coexistence equilibria is that their existence is not required for metastable patterns in which both species coexist to form and patterns outside the existence region of $(\overline{u}_1^{C,\pm}, \overline{u}_2^{C}, \overline{w}^{C,\pm})$ truly originate from a stable tree-only pattern as discussed in Section \ref{sec: Multispecies: Metastable patterns from tree patterns}. In particular, the simulation shown in Figure \ref{fig: Multispecies: Simulations: example solutions coex patterns initially tree pattern} is obtained by using parameter values for which the coexistence steady states do not exist (see the $(\beta)$ marker in Figure \ref{fig: Multispecies: Equilibria: coex ss existence and positivity plus}). The parameter region considered in this section may, however, overlap with that considered in Section \ref{sec: Multispecies: Metastable patterns from tree patterns}, and no general statement on the sizes of $A_{\min}^T$ and $A_{\min}^{C,+}$ can be made.

To gain a better understanding of the effects caused by the difference in both plant types, it is essential to understand the steady states' behaviour if the species are identical. At $\chi=0$, the coexistence steady state is 
\begin{multline}\label{eq: Multispecies: LinStab: coex steady state at chi=0}
\left(\overline{u}_1^{C,\pm}, \overline{u}_2^C, \overline{w}^{C,\pm} \right) = \\ \left(\frac{\left(A\pm\sqrt{A^2-4B_1^2} \right)}{2B_1} - \frac{b_2-B_1f}{s}, \frac{b_2-B_1f}{s}, \frac{2B_1^2}{A\pm\sqrt{A^2-4B_1^2}}\right).
\end{multline}
As remarked in Section \ref{sec: Multispecies: Model}, for $\chi=0$, the densities $u_1+u_2$ and $w$ satisfy the extended Klausmeier model. Thus, the sum $u_1+u_2$ gives rise to a continuum of steady states that satisfy
\begin{align*}
u_1+u_2 = \frac{A\pm \sqrt{A^2-4B_1^2}}{2B_1^2}, \quad \text{and} \quad w =\frac{2B_1^2}{A\pm\sqrt{A^2-4B_1^2}}.
\end{align*}
The coexistence steady state $(\overline{u}_1^{C,\pm}, \overline{u}_2^C, \overline{w}^{C,\pm})$ maps to one member of this continuum whose choice depends on the model parameters as given  by \eqref{eq: Multispecies: LinStab: coex steady state at chi=0}.

\subsection{Stability to spatially uniform perturbations}

\begin{figure}
	\centering
	\subfloat[Spatially uniform perturbations.\label{fig: Multispecies: LinStab: coex steady state max real part of eigenvalues S-4 uniform}]{\includegraphics[width=0.48\textwidth]{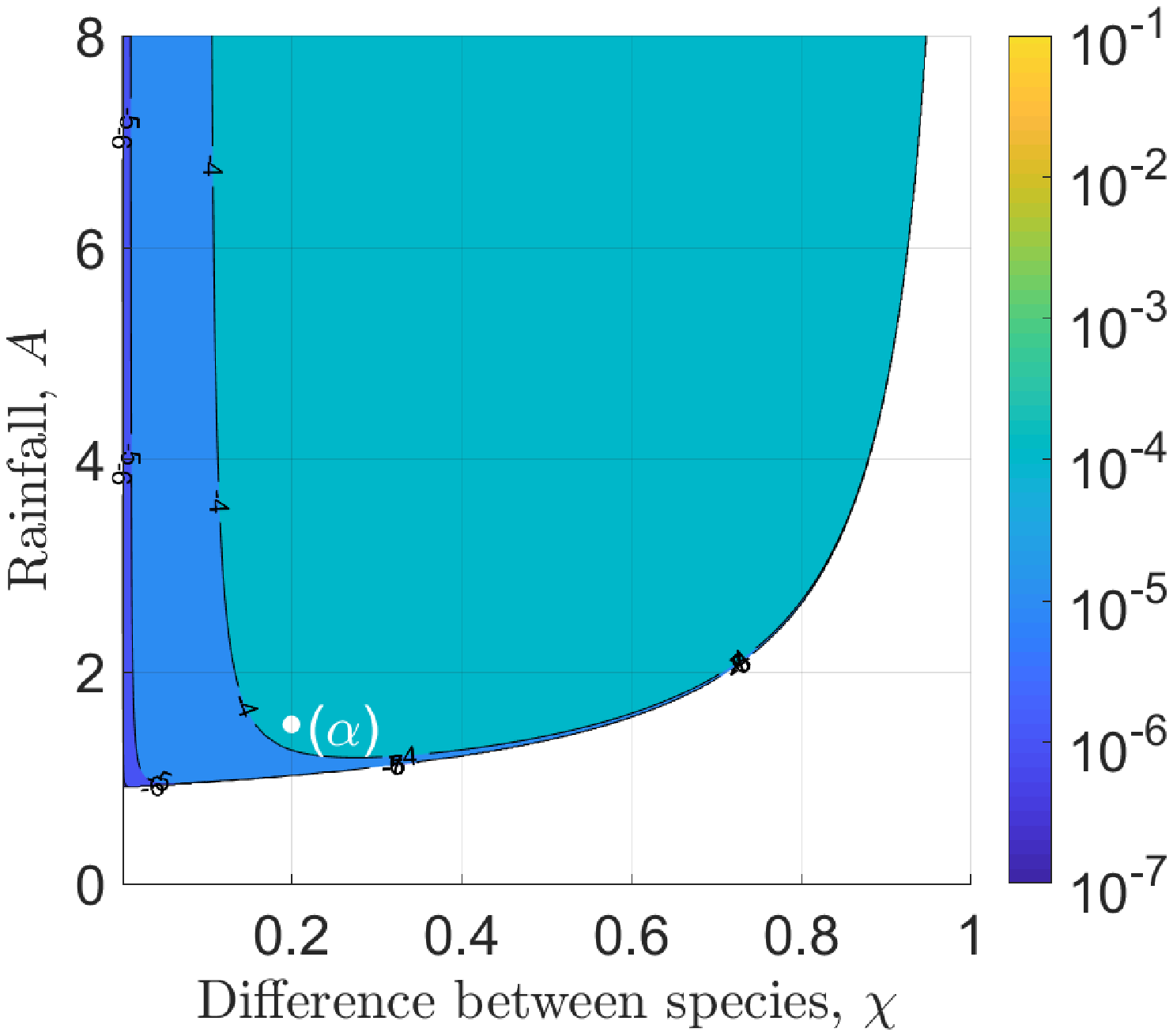}}
	\subfloat[Spatially heterogeneous perturbations.\label{fig: Multispecies: LinStab: coex steady state max real part of eigenvalues S-4 heterogeneous}]{\includegraphics[width=0.48\textwidth]{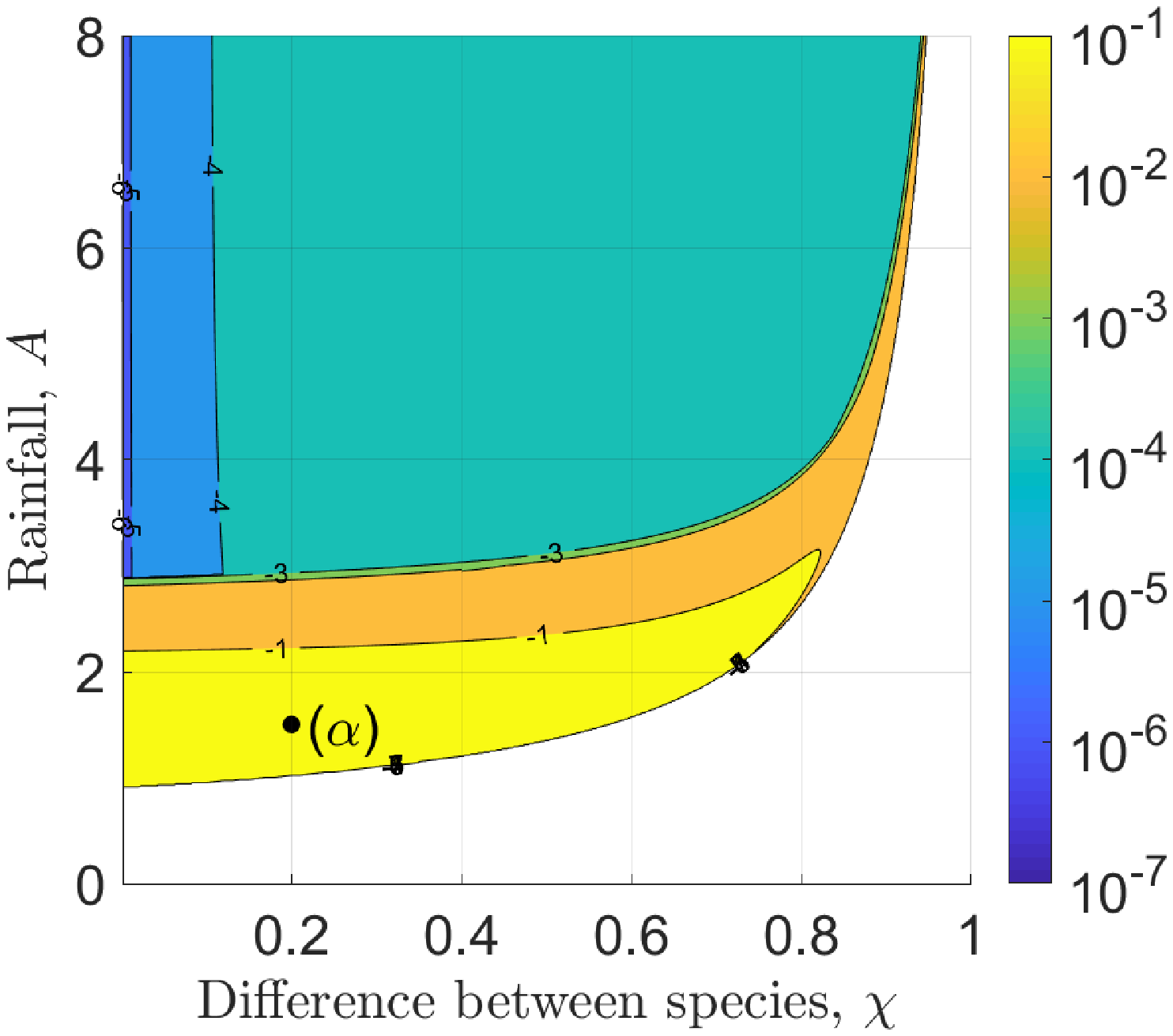}}\\
	\subfloat[Order of magnitude difference.\label{fig: Multispecies: LinStab: coex steady state max real part of eigenvalues S-4 difference}]{\includegraphics[width=0.48\textwidth]{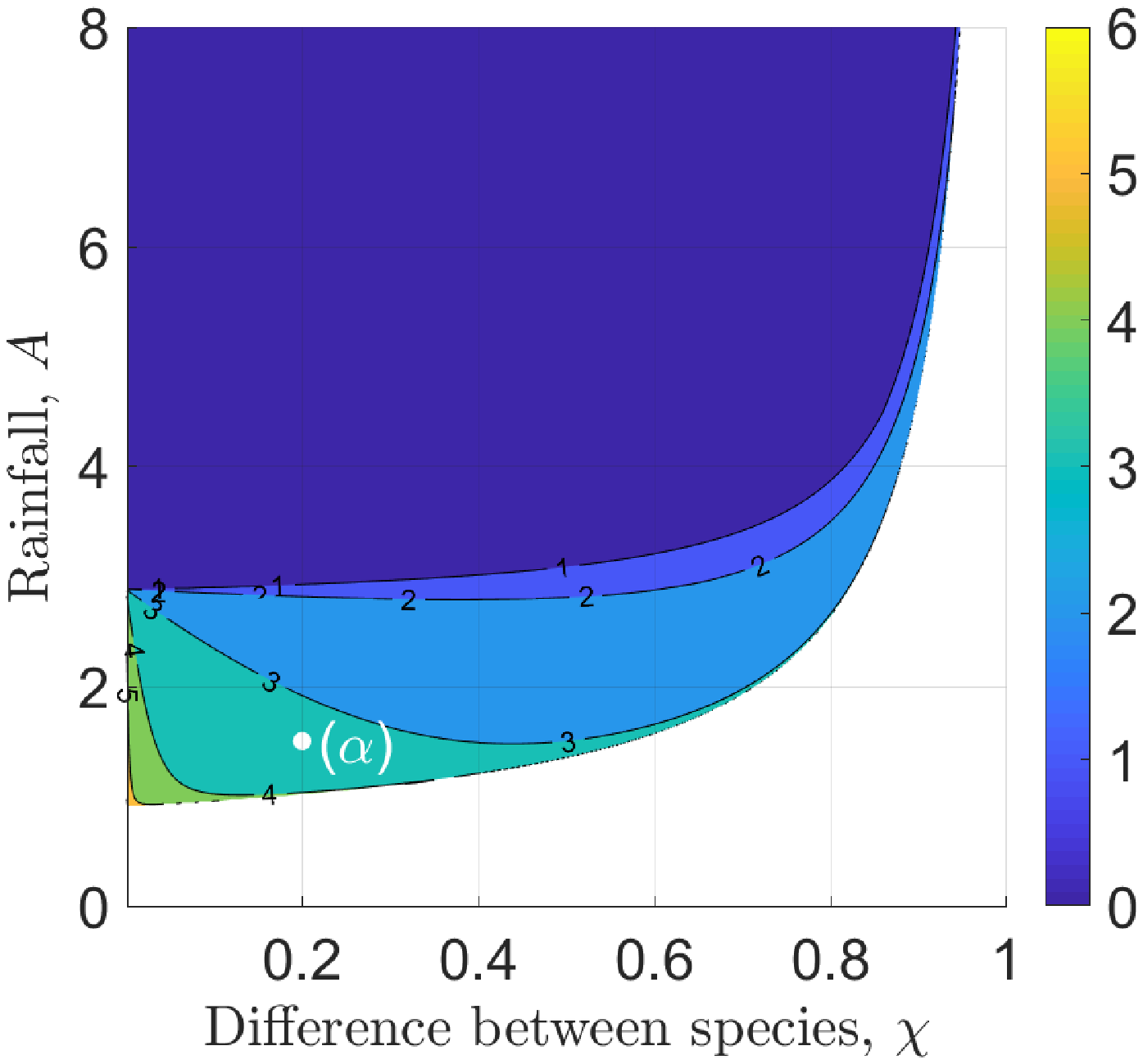}}
	
	\caption{Largest real part of eigenvalues determining stability of the coexistence steady state. Visualisation of $\max_{k>0}\{\Re(\lambda_{\cdot,1}^{C,+})\}$ in the $\chi$-$A$ parameter plane for the coexistence equilibrium $(\overline{u}_1^{C,+}, \overline{u}_2^{C}, \overline{w}^{C,+})$, where $\lambda_{\cdot,1}^{C,+}$ denotes the eigenvalue with largest real part of the Jacobian with entries \eqref{eq: Multispecies: LinStab: Jacobian} evaluated at the steady state that determine its stability to spatially uniform ((a)) and spatially heterogeneous ((b)) perturbations. The order of magnitude difference between the the results for spatially uniform and spatially heterogeneous perturbations is shown in (c). White areas indicate regions in which the steady state is negative or does not exist. The plots are obtained by evaluating $\max_{k>0}\{\Re(\lambda_{\cdot,1}^{C,+})\}$ for $0<A<8$ and $0<\chi<1$ with increments $\Delta A = 0.01$ and $\Delta \chi =0.001$. The parameters are $s=10^{-3}$, $B_1=0.45$, $b_2 = 0.0055$, $f=0.01$, $h=0.01$, $D_0=0.01$, $d=500$. The marker $(\alpha)$ refers to the parameter values used in the simulations presented in Figure \ref{fig: Multispecies: Simulations: example solutions coex patterns initally coex ss}.}\label{fig: Multispecies: LinStab: coex steady state max real part of eigenvalues}
\end{figure}

Similar to the analysis in Section \ref{sec: Multispecies: Metastable patterns from tree patterns}, linear stability analysis can be used to investigate the existence of patterns arising from the coexistence steady state $(\overline{u}_1^{C,\pm}, \overline{u}_2^C, \overline{w}^{C,\pm})$. The algebraic complexity of the Jacobian with entries \eqref{eq: Multispecies: LinStab: Jacobian} evaluated at both coexistence equilibria does not allow an analytic derivation of stability conditions similar to those for the single-species states in Section \ref{sec: Multispecies: Metastable patterns from tree patterns}. Instead, we performed a systematic numerical investigation of the Jacobian's eigenvalues $\lambda_u^{C,\pm} \in \C$ that determine the steady states' stability to spatially uniform perturbations in the respective positivity regions. This suggests that both steady states are unstable. The instability of $(\overline{u}_1^{C,+}, \overline{u}_2^{C}, \overline{w}^{C,+})$, however, is caused by an eigenvalue of small size, denoted by $\lambda_{u,1}^{C,+}$, i.e. $0<\max_{\lambda_{u}}\{\Re(\lambda_u^{C,+})\} = \Re(\lambda_{u,1}^{C,+})\ll 1$, where the maximum is taken over all eigenvalues $\lambda_u^{C,+}$ of the Jacobian $J^{C,+}=(j)_{k\ell}^{C,+}$, $k,\ell=1,2,3$ evaluated at the steady state (see Figure \ref{fig: Multispecies: LinStab: coex steady state max real part of eigenvalues S-4 uniform}). The metastability associated with the small size of $\Re(\lambda_{u,1}^{C,+})$ is, as in the case discussed in Section \ref{sec: Multispecies: Metastable patterns from tree patterns}, due to the species' similar average fitness, i.e. the small difference of $B_2-B_1F$. Indeed, an application of determinant-preserving elementary row operations shows
\begin{multline*}
\det\left(J^{C,+}\right) = \det \matthree{j_{11}^{C,+}}{j_{12}^{C,+}}{j_{13}^{C,+}}{0}{B_2-B_1F}{0}{j_{31}^{C,+}}{j_{32}^{C,+}}{j_{33}^{C,+}}  \\ = \left(B_2-B_1F \right) \left( j_{11}^{C,+}j_{33}^{C,+}-j_{13}^{C,+}j_{31}^{C,+}\right) = O\left(B_2-B_1F\right).
\end{multline*}
The equilibrium is only of biological relevance if $B_2>B_1F$. Thus, as discussed in Section \ref{sec: Multispecies: Coexistence patterns through tree patterns: Metastability}, $|B_2-B_1F| \ll 1$, and hence $|\det J| \ll 1$. Since the determinant of a matrix is the product of its eigenvalues, this shows the small size of one of the Jacobian's eigenvalues. 
If $B_2-B_1F=0$ but $S\ne 0$, then the coexistence steady state $(\overline{u}_1^{C,\pm},\overline{u}_2^{C} ,\overline{w}^{C,\pm})$ reduces to the grass-only equilibrium $(\overline{u}_1^{G,\pm},0 ,\overline{w}^{G,\pm})$ and the small eigenvalue $\lambda_{u,1}^{C,+}$ of the coexistence state corresponds to $\lambda_{u,1}^{G,\pm}$ which vanishes because $B_2-B_1F=0$. 

\subsubsection{Metastable States}

For a system initially close to the coexistence steady state $(\overline{u}_1^{C,+},\overline{u}_2^{C} ,\overline{w}^{C,+})$ the small size of the only positive real part of the Jacobian's eigenvalues leads to a slow transition away from the equilibrium in the spatially uniform setting. If spatially nonuniform perturbations of the steady state are considered, this transition occurs via metastable coexistence patterns of both species, subject to sufficiently low rainfall levels. This is quantified by linear stability analysis which shows that the maximum real part of the corresponding Jacobian's eigenvalues exceeds $\Re(\lambda_{u,1}^{C,+})$ by several orders of magnitude (see Figures \ref{fig: Multispecies: LinStab: coex steady state max real part of eigenvalues S-4 heterogeneous} and \ref{fig: Multispecies: LinStab: coex steady state max real part of eigenvalues S-4 difference} for a visualisation). In other words, $\max_{k\ge0}\{\Re(\lambda_{s,1}^{C,+}(k^2))\} \gg \Re(\lambda_{u,1}^{C,+})$, where $\lambda_{s,1}^{C,+}(k^2)$ denotes the eigenvalue of $J^{C,+} - \operatorname{diag}(k^2, Dk^2, dk^2)$ with the largest real part. This leads to a quick establishment of a coexistence pattern about the steady state from a spatially non-uniform perturbation which then persists for a long time before transiting to a stable one-species state. The growth rate that causes the formation of spatial patterns is given by
\begin{align}\label{eq: Multispecies: LinStab: coexisting ss dispersion relation}
\Re\left(\lambda_{s,1}^{C,+}\left(k^2\right) \right) = \alpha\left(k^2\right)+\Re\left(\frac{\left(\beta\left(k^2\right) + \sqrt{\gamma\left(k^2\right)} \right)^{\frac{2}{3}} + \delta\left(k^2\right)}{\left(\beta\left(k^2\right) + \sqrt{\gamma\left(k^2\right)} \right)^{\frac{1}{3}}} \right),
\end{align}
where $\alpha$, $\beta$, $\gamma$ and $\delta$ are polynomials in $k^2$. Due to the algebraic complexity of the eigenvalue,  an analytic determination of the pattern-defining features is impractical. Instead, we studied it numerically to determine the existence and possible wavelengths of a metastable pattern. 

\begin{figure}
	\centering
	\subfloat[\label{fig: Multispecies: LinStab: coex steady state dispersion relation small chi}]{\includegraphics[width=0.48\textwidth]{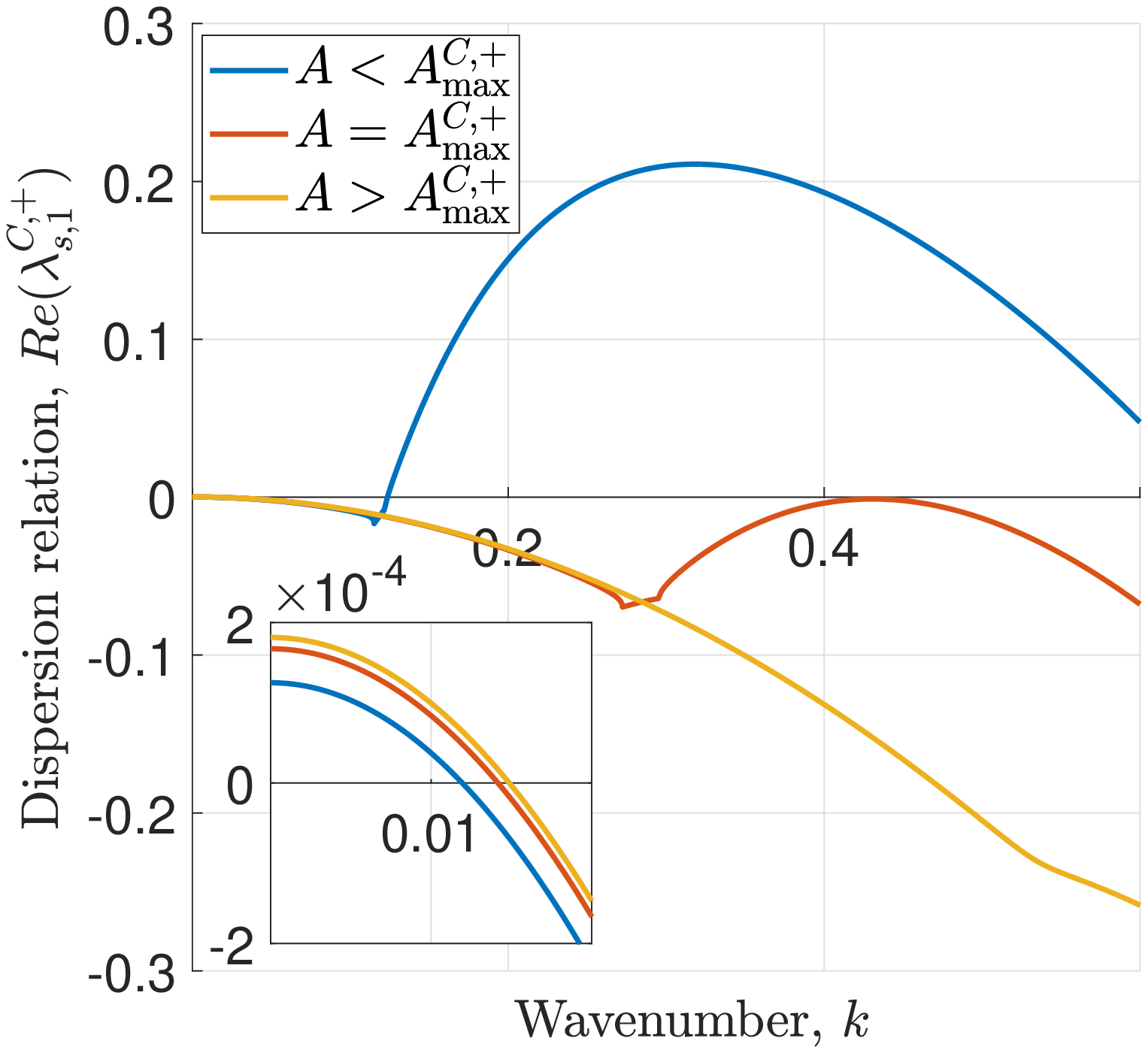}}
	\subfloat[\label{fig: Multispecies: LinStab: coex steady state dispersion relation large chi discontinuities}]{\includegraphics[width=0.48\textwidth]{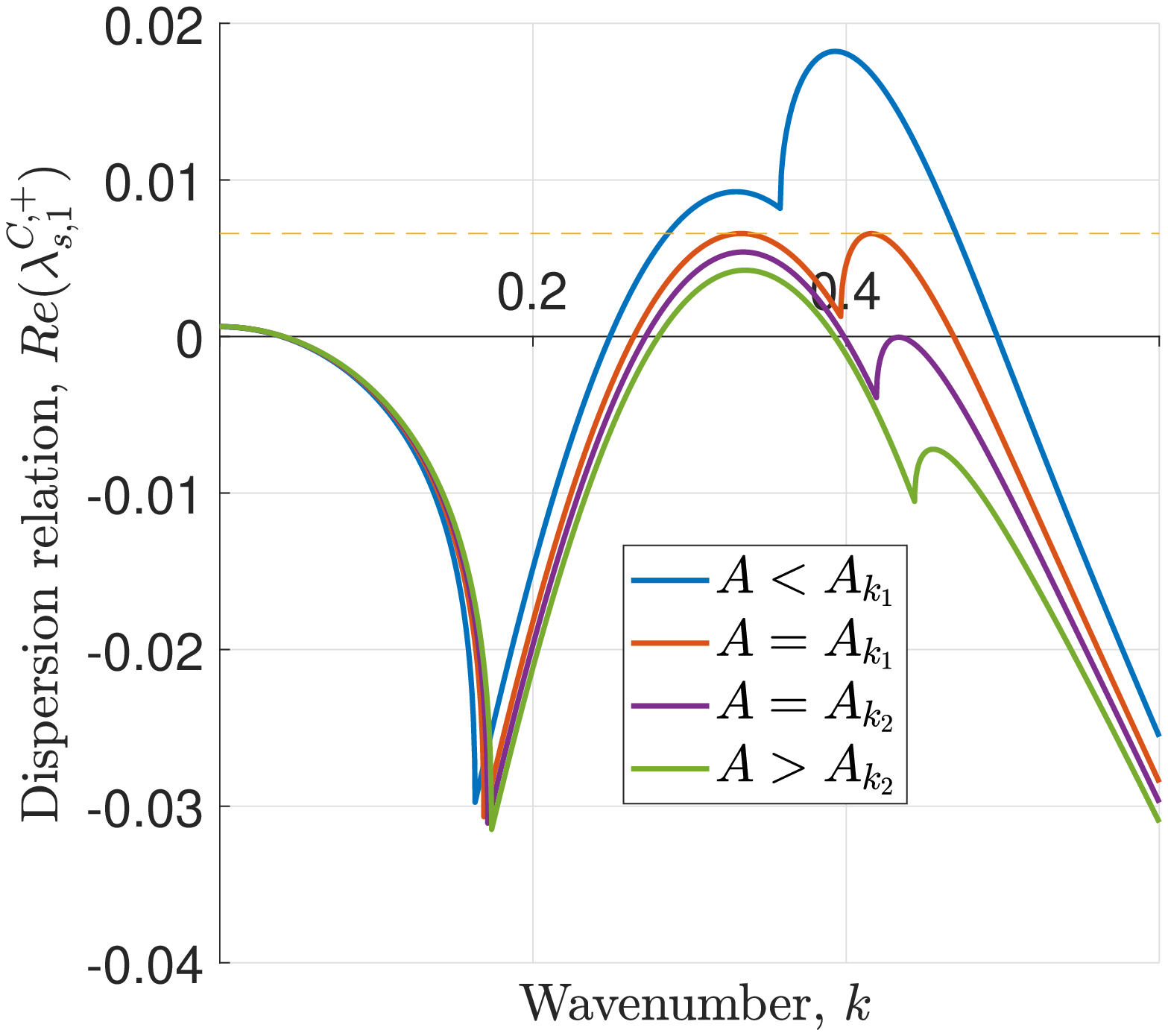}}
	\caption{Dispersion relation for patterns with species coexistence. The dispersion relation is visualised for different rainfall levels $A$ and fixed $\chi=0.2$ ((a)) and $\chi=0.86$ ((b)). The inset in (a) shows the behaviour close to the origin. The dotted line in (b) indicates the equality of the local maxima for $A=A_{k_1}$. The other parameters are $s=10^{-3}$, $B_1=0.45$, $b_2 = 0.0055$, $f=0.01$, $h=0.01$, $D_0=0.01$, $d=500$.}\label{fig: Multispecies: LinStab: coex steady state dispersion relation}
\end{figure}

As rainfall $A$ increases from the minimum $A_{\min}^{C,+}$, $\max_{k\ge0}\{\Re(\lambda_{s,1}^{C,+}(k^2))\}$ decreases and there exists a critical value of precipitation $A_{\max}^{C,+}$ beyond which $\max_{k\ge0}\{\Re(\lambda_{s,1}^{C,+}(A;k^2))\}=\Re(\lambda_{u,1}^{C,+}(A))$ (Figure \ref{fig: Multispecies: LinStab: coex steady state dispersion relation small chi}). In particular, there is a discontinuity in $k_{\max}^{C,+}:=\arg\max_{k\ge0}\{\Re(\lambda_{s,1}^{C,+}(k^2))\}$ at $A=A_{\max}^{C,+}$, because the maximum real part of the eigenvalues attains its maximum at $k=0$ for $A>A_{\max}^{C,+}$, but $k_{\max}^{C,+} \nrightarrow 0$ as $A \uparrow A_{\max}^{C,+}$. This threshold is an upper bound for the existence of metastable coexistence patterns and is visualised in Figure \ref{fig: Multispecies: LinStab: coex steady state max real part of eigenvalues S-4 difference}. For rainfall levels above this threshold, metastable coexistence of both plant species still occurs, albeit not as a pattern. Spatial heterogeneity does not cause the formation of patterns in this case as $\Re(\lambda_{s,1}^{C,+}(A))$ attains its maximum at $k=0$. The small size of $\Re(\lambda_{u,1}^{C,+})$ still causes a solution slightly perturbed from the coexistence steady state to remain close to the equilibrium for a long time. This gives rise to a metastable state in which both vegetation types are present uniformly in space.

\subsubsection{Wavelength}\label{sec: Multispecies: LinStab: wavelength coex}

\begin{figure}
	\centering
	\subfloat[Minimum wavelength.\label{fig: Multispecies: LinStab: coex steady state wavelengths of patterns min}]{\includegraphics[width=0.48\textwidth]{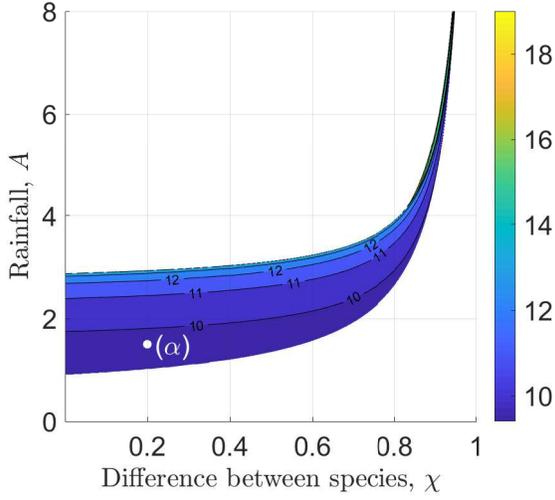}}
	\subfloat[Maximum wavelength.\label{fig: Multispecies: LinStab: coex steady state wavelengths of patterns max}]{\includegraphics[width=0.48\textwidth]{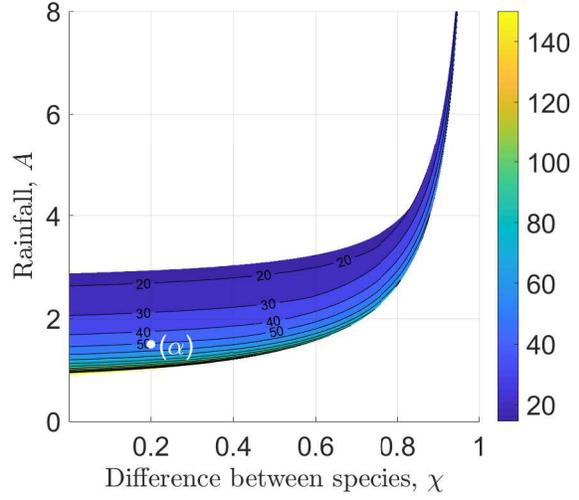}}\\
	\subfloat[Wavelength arising from most unstable wavenumber.\label{fig: Multispecies: LinStab: coex steady state wavelengths of patterns most unstable}]{\includegraphics[width=0.48\textwidth]{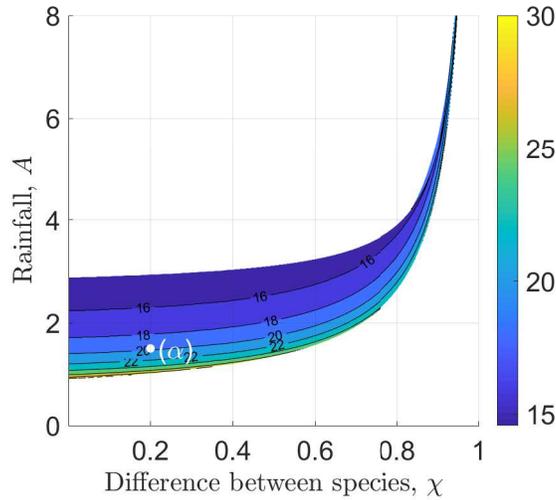}}
	\caption{Wavelength of metastable coexistence patterns. This figure visualises contours of the wavelength associated with the wavenumber yielding the largest growth (c) as well as lower (a) and upper bounds (b) that arise from linear stability analysis. For details on the creation of the plots and the parameter values see Figure \ref{fig: Multispecies: LinStab: coex steady state max real part of eigenvalues}. For a comparison to the wavelengths of the single-species pattern see Figure \ref{fig: Multispecies: LinStab: single species pattern wavelengths}.}\label{fig: Multispecies: LinStab: coex steady state wavelengths of patterns}
\end{figure}

Linear stability analysis further provides an insight into the wavelength of patterns. Typically the wavelength of a pattern is dominated by the wavenumber yielding the largest growth. However, since the wavelength of a pattern is an inherently nonlinear property different modes may be selected due to effects such as hysteresis. In this case the roots of $\Re(\lambda_{s,1}^{C,+}(k^2))$ provide an upper and lower bound for the wavelength. The numerical investigation of the dispersion relation shows that pattern wavelength increases with decreasing rainfall, in line with results shown in Section \ref{sec: Multispecies: Metastable patterns from tree patterns} and previous results on the single-species Klausmeier model on sloped ground \cite{Sherratt2005, Sherratt2013IV}. In other words, the distance between vegetation patches is larger in regions in which a smaller amount of the limiting water resource is available. An increase in the difference between the two plant species also causes an increase in the wavelength difference, but this increase is small compared to changes caused by precipitation fluctuations. A visualisation of the wavelength is given in Figure \ref{fig: Multispecies: LinStab: coex steady state wavelengths of patterns}.
A further complication in the calculation of the wavelength through linear stability analysis arises through the algebraic complexity of the dispersion relation \eqref{eq: Multispecies: LinStab: coexisting ss dispersion relation} which causes a discontinuity in the most unstable mode and hence also the largest root in a subset of the parameter space considered in this analysis. The discontinuities arise from the existence of two local maxima of $\Re(\lambda_{s,1}^{C,+}(k^2))$, one of which occurs for $k_1<k<k_2$, which is the positivity region of $\gamma(k^2)$, while the other local maximum is attained for $k>k_2$. Consequently, there exists a critical value of the precipitation parameter $A_{k_1}$ at which there is a discontinuity in $\arg\max_{k\ge0}\Re(\lambda_{s,1}^{C,+}(k^2))$ because both local maxima coincide (see Figure \ref{fig: Multispecies: LinStab: coex steady state dispersion relation large chi discontinuities}). 
Similarly, the rainfall value $A_{k_2}$ at which $\max_{k\ge k_2}\Re(\lambda_{s,1}^{C,+}(k^2)) = 0$, causes a discontinuity in the largest root of the dispersion relation and thus in the lower bound for the wavelength of the coexistence pattern.


\section{Discussion}\label{sec: Multispecies: Discussion}

Our work predicts that coexistence of two plant species competing for the same limiting resource can occur as a long transient state, even if coexistence is inherently unstable. Such a metastable behaviour is characterised by the small size of the only positive eigenvalue of the equilibrium from which the coexistence arises. Coexistence of two species in such a metastable state is enabled by a balance of both species' average fitness which is measured by the ratio of a species' capability to convert water into new biomass to its mortality rate. In the nondimensional model parameters this balance corresponds to the small size of $B_2-B_1F$, the quantity that controls the size of the eigenvalue causing the instability. 

In ecology, the understanding of transient states is of utmost importance as many ecosystems never reach an equilibrium state. Disturbances such as changes to grazing patterns or climate change interrupt the convergence to a steady state on a frequent basis, and thus keep systems in perpetual transients \cite{Sprugel1991,Svenning2013}. The occurrence of such disequilibrium states is not specific to savanna and dryland biomes but also occurs in ecosystems of other climate zones \cite{Serra-Diaz2018}. While we have not investigated the system's response to changes in environmental conditions, such as variability in precipitation or a changes in water evaporation due to temperature fluctuations, the analysis presented in this paper can provide an insight into the dynamics of such transient states by investigating their origin, fate and some of their properties. 

We have established two possible origins of metastable states in the multispecies model \eqref{eq: Multispecies: Model: nondimensional model}: a spatially uniform equilibrium in which both species coexist (Section \ref{sec: Multispecies: Metastable patterns from coexistence steady state}) and a one-species tree pattern that is unstable to the introduction of the herbaceous species (Section \ref{sec: Multispecies: Metastable patterns from tree patterns}). For the latter, the consideration of the interspecific shading feedback is not necessary. The direct interspecific competition does, however, cause a further decrease in the unstable eigenvalue \eqref{eq: Multispecies: LinStab: unstable eigenvalue tree to grass introduction}, by further reducing the average fitness difference between both species. Large shading effects may also tip that balance in favour of the tree species, stabilising the tree pattern and thus preventing the formation of a metastable coexistence pattern from an invasion-type scenario (see Figure \ref{fig: Multispecies: LinStab: stability diagram change all parameters}).

On the other hand, the inclusion of the shading effect is essential for the existence of metastable states arising from a coexistence equilibrium as a direct interspecific competition term is necessary for the existence of such a steady state. Coexistence at equilibrium without the presence of a shading effect is only possible if the average fitness of both species are equal, i.e. $B_2=B_1F$, a highly unlikely scenario unless both species are the same. Similar to a previous analysis of a multi-species model in dryland ecosystems by \cite{Nathan2013} we did not consider this special case as it lacks biological relevance. Nevertheless, the lack of a shading feedback does not necessarily prevent the establishment of a coexistence pattern from perturbations to a spatially uniform configuration of both species similar to that visualised in Figure \ref{fig: Multispecies: Simulations: example solutions coex patterns initally coex ss}. If the species differ in their dispersal behaviour, the faster dispersing species can establish a spatial pattern (provided precipitation is sufficiently low) and can act as an ecosystem engineer by redistributing the water resource to which the slow disperser can adapt and form a pattern itself. As discussed in slightly different settings by \cite{Nathan2013} and \cite{Baudena2013}, this supports the existence of coexistence patterns. In particular, this pushes the system into a state to which the theory presented in Section \ref{sec: Multispecies: Metastable patterns from tree patterns} can be applied. Hence, if one of the two corresponding single-species states is unstable to the introduction of the competitor via a very small eigenvalue, the system remains in the coexistence pattern state for a long time. This observation emphasises the difficulty of inferring the origin of a metastable multi-species patterned state, which is beyond the scope of this paper.

The wavelength of the pattern may provide a useful tool in predicting the fate of a coexistence pattern, but potential shortfalls (linearisation, neglection of hysteresis effects) in the determination of the wavelength need to be taken into account. Our analysis of the patterns' wavelengths shows that the wavelength of a single-species tree pattern (Figure \ref{fig: Multispecies: LinStab: single species pattern wavelengths}) is very similar to that of a pattern in which the tree species coexists with the grass species (Figure \ref{fig: Multispecies: LinStab: coex steady state wavelengths of patterns}). However, if both species differ significantly (the parameter $\chi$ being close to unity), linear stability analysis predicts single-species grass patterns at a smaller wavelength than coexistence patterns. Thus, if a pattern in which both species coexist occurs at an atypical mode that differs from the results presented in Sections \ref{sec: Multispecies: LinStab: wavelength single} and \ref{sec: Multispecies: LinStab: wavelength coex} and better fits the wavelength prediction of a one-species pattern (such as in the later stages of the solution visualised in Figure \ref{fig: Multispecies: Simulations: example solutions coex patterns initially tree pattern}), it can be concluded that the metastable pattern eventually reduces to a one-species pattern to which the observed wavelength corresponds.

We have restricted our analysis in this paper to the two-species model \eqref{eq: Multispecies: Model: nondimensional model} to focus on the analytical investigation of pattern existence. Numerical simulations of a three-species model similar to \eqref{eq: Multispecies: Model: general multispecies no shading} with $n=3$, but with the addition of multiple, hierarchical interspecific interaction terms, also yield metastable patterned solutions in which all three species coexist, provided their average fitness differences are small. Coexistence through metastability can further occur for just a subset of all species in the model. Indeed, our numerical experiments show that if one of the species has a lower average fitness, then the community of superior species outcompetes the inferior species on a short timescale and forms a metastable coexistence state in which it remains on a long timescale. We thus hypothesise that the metastability property discussed in this paper is not specific to the two-species model \eqref{eq: Multispecies: Model: nondimensional model} but can be extended to a larger community of plant species in desert ecosystems. Moreover, our simulations of the three-species model indicate that the crucial condition for the existence of metastable solutions - small average fitness differences between species - is carried over to systems of more diverse plant communities.

The concept of a metastable solution to a system is not new. Metastability has, for example, been studied in the Cahn-Hillard equation \cite{Bates1994,Bates1995}, in chemotactic models \cite{Potapov2005} and microwave heating models \cite{Iron2004}. The occurrence of a slow transient between two stable states has even been briefly commented on in the analysis of a more complex multi-species model of desert plants \cite{Gilad2007a}, without the attempt to provide a detailed investigation of the phenomenon. It is worth emphasising that we characterise metastability by the small size of the only positive eigenvalue of an equilibrium. In landscape ecology, however, the term metastability usually has a broader meaning as it describes a stable system whose single components are changing over time due to disturbance and recovery effects \cite{Zimmerman2010}.

The model in this paper is based on the Klausmeier model \cite{Klausmeier1999}, which deliberately reduces the description of the dynamics responsible for the formation of vegetation patterns in arid environments to the infiltration feedback arising from a soil modification caused by plants. A range of more complex models exist (see \cite{Zelnik2013} for a review of the most commonly used models) that capture a number of additional features of dryland ecosystems, such as nonlocal plant dispersal \cite{Baudena2013,Pueyo2008,Pueyo2010,Eigentler2018nonlocalKlausmeier, Alfaro2018},  different dynamics of soil and surface water \cite{Rietkerk2002,HilleRisLambers2001}, nonlocal water uptake due to extended root networks \cite{Gilad2004}, more realistic grazing/browsing effects \cite{Siero2018, Siero2019} or autotoxicity \cite{Marasco2014}. Simulation-based approaches have to some extent addressed the influence of these feedbacks on the coexistence of species \cite{Kyriazopoulos2014,Gilad2007a}, but an analytical approach similar to that presented in this paper may provide further insight into the way in which these additional assumptions affect coexistence mechanisms.

A natural extension of the work presented in this paper would be an investigation of the metastability property in a two-dimensional space domain. The linear stability analysis from Sections \ref{sec: Multispecies: Metastable patterns from tree patterns} and \ref{sec: Multispecies: Metastable patterns from coexistence steady state} can be carried over to a higher space dimension, but does not provide any new information on the metastable behaviour of a patterned solution. Instead, numerical simulations could provide more insights into the coexistence pattern's properties away from the Turing bifurcation locus, such as a classification of its type (gap pattern, labyrinth pattern, stripe pattern or spot pattern) along the precipitation gradient \cite{Meron2012}. The combination of adding an additional space dimension with the long runtimes required to capture the metastable behaviour of the system would, however, incur a significant computational cost.

A final area of potential future work concerns variabilities in environmental conditions, which have not been addressed in this paper. Effects such as rainfall seasonality \cite{Guttal2007,Kletter2009,Baudena2007}, rainfall intermittency \cite{Baudena2007,Kletter2009,Ursino2006,Siteur2014a}, periodic variation in precipitation \cite{Tzuk2019} or topographic heterogeneity \cite{Gandhi2018} are known to be significant for vegetation patterns and have been studied using single-species models. It could therefore be of interest to extend those approaches to multi-species ecosystems to develop an understanding of how such heterogeneities affect the coexistence of species and, in particular, the metastability property of the model presented in this paper. Indeed, simulations of our multispecies model under seasonal precipitation regimes suggest that rainfall seasons of intermediate length (150 - 250 days per year) prolong the time the system remains in a coexistence state. Initial simulations, however, also suggest that inherently nonlinear properties such as pattern wavelength have a significant effect on the system's transient behaviour under temporal variations of environmental conditions. A detailed investigation of this phenomenon is therefore beyond the scope of this paper, but would present new valuable insights into coexistence of plant species in dryland ecosystems.

\section*{Acknowledgements}
Lukas Eigentler was supported by The Maxwell Institute Graduate School in Analysis and its Applications, a Centre for Doctoral Training funded by the UK Engineering and Physical Sciences Research Council (grant EP/L016508/01), the Scottish Funding Council, Heriot-Watt University and the University of Edinburgh.

\clearpage
\bibliography{bibliography.bib}

\begin{thebibliography}{10}
\expandafter\ifx\csname url\endcsname\relax
  \def\url#1{\texttt{#1}}\fi
\expandafter\ifx\csname doi\endcsname\relax
  \def\doi#1{\burlalt{doi:#1}{http://dx.doi.org/#1}}\fi
\expandafter\ifx\csname urlprefix\endcsname\relax\def\urlprefix{URL }\fi
\expandafter\ifx\csname href\endcsname\relax
  \def\href#1#2{#2}\fi
\expandafter\ifx\csname burlalt\endcsname\relax
  \def\burlalt#1#2{\href{#2}{#1}}\fi

\bibitem{Alfaro2018}
M.~Alfaro, H.~Izuhara, and M.~Mimura.
\newblock On a nonlocal system for vegetation in drylands.
\newblock {\em J. Math. Biol.}, 77(6-7):1761--1793, 2018.
\newblock \doi{10.1007/s00285-018-1215-0}.

\bibitem{Bastiaansen2018}
R.~Bastiaansen, O.~Jaïbi, V.~Deblauwe, M.~B. Eppinga, K.~Siteur, E.~Siero,
  S.~Mermoz, A.~Bouvet, A.~Doelman, and M.~Rietkerk.
\newblock Multistability of model and real dryland ecosystems through spatial
  self-organization.
\newblock {\em Proceedings of the National Academy of Sciences}, pages
  11256--11261, 2018.
\newblock \doi{10.1073/pnas.1804771115}.

\bibitem{Bates1994}
P.~Bates and J.~Xun.
\newblock Metastable patterns for the {C}ahn-{H}illiard equation, part {I}.
\newblock {\em Journal of Differential Equations}, 111(2):421--457, 1994.
\newblock \doi{10.1006/jdeq.1994.1089}.

\bibitem{Bates1995}
P.~Bates and J.~Xun.
\newblock Metastable patterns for the {C}ahn-{H}illiard equation: Part {II}.
  layer dynamics and slow invariant manifold.
\newblock {\em Journal of Differential Equations}, 117(1):165--216, 1995.
\newblock \doi{10.1006/jdeq.1995.1052}.

\bibitem{Baudena2007}
M.~Baudena, G.~Boni, L.~Ferraris, J.~von Hardenberg, and A.~Provenzale.
\newblock Vegetation response to rainfall intermittency in drylands: Results
  from a simple ecohydrological box model.
\newblock {\em Adv. Water Resour.}, 30(5):1320 -- 1328, 2007.
\newblock \doi{10.1016/j.advwatres.2006.11.006}.

\bibitem{Baudena2010}
M.~Baudena, F.~D’Andrea, and A.~Provenzale.
\newblock An idealized model for tree–grass coexistence in savannas: the role
  of life stage structure and fire disturbances.
\newblock {\em J. Ecol.}, 98(1):74--80, 2010.
\newblock \doi{10.1111/j.1365-2745.2009.01588.x}.

\bibitem{Baudena2013}
M.~Baudena and M.~Rietkerk.
\newblock Complexity and coexistence in a simple spatial model for arid savanna
  ecosystems.
\newblock {\em Theor. Ecol.}, 6(2):131--141, 2013.
\newblock \doi{10.1007/s12080-012-0165-1}.

\bibitem{Bennett2019}
J.~J.~R. Bennett and J.~A. Sherratt.
\newblock Long-distance seed dispersal affects the resilience of banded
  vegetation patterns in semi-deserts.
\newblock {\em J. Theor. Biol.}, 481:151--161, 2018.
\newblock \doi{10.1016/j.jtbi.2018.10.002}.

\bibitem{Borgogno2009}
F.~Borgogno, P.~D'Odorico, F.~Laio, and L.~Ridolfi.
\newblock Mathematical models of vegetation pattern formation in ecohydrology.
\newblock {\em Rev. Geophys.}, 47:RG1005, 2009.
\newblock \doi{10.1029/2007RG000256}.

\bibitem{Buis2009}
E.~Buis, A.~Veldkamp, B.~Boeken, and N.~{van Breemen}.
\newblock Controls on plant functional surface cover types along a
  precipitation gradient in the {N}egev {D}esert of {I}srael.
\newblock {\em J. Arid. Environ.}, 73(1):82 -- 90, 2009.
\newblock \doi{10.1016/j.jaridenv.2008.09.008}.

\bibitem{Callegaro2018}
C.~Callegaro and N.~Ursino.
\newblock Connectivity of niches of adaptation affects vegetation structure and
  density in self-organized (dis-connected) vegetation patterns.
\newblock {\em Land Degradation \& Development}, 29(8):2589--2594, 2018.
\newblock \doi{10.1002/ldr.2759}.

\bibitem{Consolo2019}
G.~Consolo, C.~Curr{\`{o}}, and G.~Valenti.
\newblock Supercritical and subcritical {T}uring pattern formation in a
  hyperbolic vegetation model for flat arid environments.
\newblock {\em Physica D}, 398:141--163, 2019.
\newblock \doi{10.1016/j.physd.2019.03.006}.

\bibitem{Cornet1988}
A.~Cornet, J.~Delhoume, and C.~Monta{\~{n}}a.
\newblock {\em Dynamics of striped vegetation patterns and water balance in the
  {C}hihuahuan {D}esert}, pages 221--231.
\newblock SPB Academic Publishing, The Hague, 1988.

\bibitem{Corrado2014}
R.~Corrado, A.~M. Cherubini, and C.~Pennetta.
\newblock Early warning signals of desertification transitions in semiarid
  ecosystems.
\newblock {\em Phys. Rev. E: Stat., Nonlinear, Soft Matter Phys.}, 90:062705,
  2014.
\newblock \doi{10.1103/PhysRevE.90.062705}.

\bibitem{Dakos2011}
V.~Dakos, S.~K{\'{e}}fi, M.~Rietkerk, E.~H. {van Nes}, and M.~Scheffer.
\newblock Slowing down in spatially patterned ecosystems at the brink of
  collapse.
\newblock {\em Am. Nat.}, 177(6):E153--E166, 2011.
\newblock \doi{10.1086/659945}.

\bibitem{Deblauwe2008}
V.~Deblauwe, N.~Barbier, P.~Couteron, O.~Lejeune, and J.~Bogaert.
\newblock The global biogeography of semi-arid periodic vegetation patterns.
\newblock {\em Global Ecol. Biogeogr.}, 17(6):715--723, 2008.
\newblock \doi{10.1111/j.1466-8238.2008.00413.x}.

\bibitem{Deblauwe2012}
V.~Deblauwe, P.~Couteron, J.~Bogaert, and N.~Barbier.
\newblock Determinants and dynamics of banded vegetation pattern migration in
  arid climates.
\newblock {\em Ecol. Monogr.}, 82(1):3--21, 2012.
\newblock \doi{10.1890/11-0362.1}.

\bibitem{Herbes2001}
J.-M. d'Herb{\`e}s, C.~Valentin, D.~J. Tongway, and J.-C. Leprun.
\newblock {\em Banded Vegetation Patterns and Related Structures}, pages 1--19.
\newblock Springer New York, New York, NY, 2001.
\newblock \doi{10.1007/978-1-4613-0207-0_1}.

\bibitem{Dickovick2014}
J.~T. Dickovick.
\newblock {\em Africa 2014-2015}.
\newblock World Today (Stryker). Rowman \& Littlefield Publishers, 2014.

\bibitem{DOnofrio2015}
D.~D'Onofrio, M.~Baudena, F.~D'Andrea, M.~Rietkerk, and A.~Provenzale.
\newblock Tree-grass competition for soil water in arid and semiarid savannas:
  The role of rainfall intermittency.
\newblock {\em Water Resour. Res.}, 51(1):169--181, 2015.
\newblock \doi{10.1002/2014WR015515}.

\bibitem{Dunkerley2002}
D.~Dunkerley and K.~Brown.
\newblock Oblique vegetation banding in the {A}ustralian arid zone:
  implications for theories of pattern evolution and maintenance.
\newblock {\em J. Arid. Environ.}, 51(2):163 -- 181, 2002.
\newblock \doi{10.1006/jare.2001.0940}.

\bibitem{Eigentler2018nonlocalKlausmeier}
L.~Eigentler and J.~A. Sherratt.
\newblock Analysis of a model for banded vegetation patterns in semi-arid
  environments with nonlocal dispersal.
\newblock {\em J. Math. Biol.}, 77(3):739--763, 2018.
\newblock \doi{10.1007/s00285-018-1233-y}.

\bibitem{Eldridge2000}
D.~Eldridge, E.~Zaady, and M.~Shachak.
\newblock Infiltration through three contrasting biological soil crusts in
  patterned landscapes in the {N}egev, {I}srael.
\newblock {\em CATENA}, 40(3):323 -- 336, 2000.
\newblock \doi{10.1016/S0341-8162(00)00082-5}.

\bibitem{Gandhi2018}
P.~Gandhi, L.~Werner, S.~Iams, K.~Gowda, and M.~Silber.
\newblock A topographic mechanism for arcing of dryland vegetation bands.
\newblock {\em Journal of The Royal Society Interface}, 15(147):20180508, 2018.
\newblock \doi{10.1098/rsif.2018.0508}.

\bibitem{Gilad2007a}
E.~Gilad, M.~Shachak, and E.~Meron.
\newblock Dynamics and spatial organization of plant communities in
  water-limited systems.
\newblock {\em Theor. Popul. Biol.}, 72(2):214--230, 2007.
\newblock \doi{10.1016/j.tpb.2007.05.002}.

\bibitem{Gilad2004}
E.~Gilad, J.~von Hardenberg, A.~Provenzale, M.~Shachak, and E.~Meron.
\newblock Ecosystem engineers: From pattern formation to habitat creation.
\newblock {\em Phys. Rev. Lett.}, 93:098105, 2004.
\newblock \doi{10.1103/PhysRevLett.93.098105}.

\bibitem{Gilad2007}
E.~Gilad, J.~von Hardenberg, A.~Provenzale, M.~Shachak, and E.~Meron.
\newblock A mathematical model of plants as ecosystem engineers.
\newblock {\em J. Theor. Biol.}, 244(4):680 -- 691, 2007.
\newblock \doi{j.jtbi.2006.08.006}.

\bibitem{Gowda2016}
K.~Gowda, Y.~Chen, S.~Iams, and M.~Silber.
\newblock Assessing the robustness of spatial pattern sequences in a dryland
  vegetation model.
\newblock {\em Proc. R. Soc. Lond. A}, 472:20150893, 2016.
\newblock \doi{10.1098/rspa.2015.0893}.

\bibitem{Gowda2018}
K.~Gowda, S.~Iams, and M.~Silber.
\newblock Signatures of human impact on self-organized vegetation in the {H}orn
  of {A}frica.
\newblock {\em Sci. Rep.}, 8:1--8, 2018.
\newblock \doi{10.1038/s41598-018-22075-5}.

\bibitem{Guttal2007}
V.~Guttal and C.~Jayaprakash.
\newblock Self-organization and productivity in semi-arid ecosystems:
  Implications of seasonality in rainfall.
\newblock {\em J. Theor. Biol.}, 248(3):490 -- 500, 2007.
\newblock \doi{10.1016/j.jtbi.2007.05.020}.

\bibitem{Hemming1965}
C.~F. Hemming.
\newblock Vegetation arcs in {S}omaliland.
\newblock {\em J. Ecol.}, 53(1):57--67, 1965.
\newblock \doi{10.2307/2257565}.

\bibitem{HilleRisLambers2001}
R.~HilleRisLambers, M.~Rietkerk, F.~{van den Bosch}, H.~H.~T. Prins, and H.~{de
  Kroon}.
\newblock Vegetation pattern formation in semi-arid grazing systems.
\newblock {\em Ecology}, 82(1):50--61, 2001.
\newblock \doi{10.2307/2680085}.

\bibitem{Iron2004}
D.~Iron and M.~J. Ward.
\newblock {The} {stability} {and} {dynamics} {of} {hot}-{spot} {solutions} {to}
  {two} {one}-{dimensional} {microwave} {heating} {models}.
\newblock {\em Analysis and Applications}, 02(01):21--70, 2004.
\newblock \doi{10.1142/s0219530504000291}.

\bibitem{Kealy2012}
B.~J. Kealy and D.~J. Wollkind.
\newblock A nonlinear stability analysis of vegetative {T}uring pattern
  formation for an interaction--diffusion plant-surface water model system in
  an arid flat environment.
\newblock {\em Bull. Math. Biol.}, 74(4):803--833, 2012.
\newblock \doi{10.1007/s11538-011-9688-7}.

\bibitem{Klausmeier1999}
C.~A. Klausmeier.
\newblock Regular and irregular patterns in semiarid vegetation.
\newblock {\em Science}, 284(5421):1826--1828, 1999.
\newblock \doi{10.1126/science.284.5421.1826}.

\bibitem{Kletter2009}
A.~Kletter, J.~{von Hardenberg}, E.~Meron, and A.~Provenzale.
\newblock Patterned vegetation and rainfall intermittency.
\newblock {\em J. Theor. Biol.}, 256(4):574 -- 583, 2009.
\newblock \doi{10.1016/j.jtbi.2008.10.020}.

\bibitem{Kyriazopoulos2014}
P.~Kyriazopoulos, J.~Nathan, and E.~Meron.
\newblock Species coexistence by front pinning.
\newblock {\em Ecol. Complexity}, 20:271--281, 2014.
\newblock \doi{10.1016/j.ecocom.2014.05.001}.

\bibitem{Kefi2007}
S.~Kéfi, M.~Rietkerk, C.~L. Alados, Y.~Pueyo, V.~Papanastasis, A.~ElAich, and
  P.~{de Ruiter}.
\newblock Spatial vegetation patterns and imminent desertification in
  {M}editerranean arid ecosystems.
\newblock {\em Nature}, 449(7159):213--217, 2007.
\newblock \doi{10.1038/nature06111}.

\bibitem{Heras2012}
M.~M. las Heras, P.~M. Saco, G.~R. Willgoose, and D.~J. Tongway.
\newblock Variations in hydrological connectivity of {A}ustralian semiarid
  landscapes indicate abrupt changes in rainfall‐use efficiency of
  vegetation.
\newblock {\em J. Geophys. Res., G: Biogeosci.}, 117:G03009, 2012.
\newblock \doi{10.1029/2011JG001839}.

\bibitem{Marasco2014}
A.~Marasco, A.~Iuorio, F.~Carteni, G.~Bonanomi, D.~M. Tartakovsky,
  S.~Mazzoleni, and F.~Giannino.
\newblock Vegetation pattern formation due to interactions between water
  availability and toxicity in plant{\textendash}soil feedback.
\newblock {\em Bull. Math. Biol.}, 76(11):2866--2883, 2014.
\newblock \doi{10.1007/s11538-014-0036-6}.

\bibitem{Meron2012}
E.~Meron.
\newblock Pattern-formation approach to modelling spatially extended
  ecosystems.
\newblock {\em Ecol. Model.}, 234:70 -- 82, 2012.
\newblock \doi{10.1016/j.ecolmodel.2011.05.035}.
\newblock Modelling clonal plant growth: From Ecological concepts to
  Mathematics.

\bibitem{Meron2016}
E.~Meron.
\newblock Pattern formation - a missing link in the study of ecosystem response
  to environmental changes.
\newblock {\em Math. Biosci.}, 271:1--18, 2016.
\newblock \doi{10.1016/j.mbs.2015.10.015}.

\bibitem{Meron2018}
E.~Meron.
\newblock From patterns to function in living systems: Dryland ecosystems as a
  case study.
\newblock {\em Annu. Rev. Condens. Matter Phys.}, 9(1):79--103, 2018.
\newblock \doi{10.1146/annurev-conmatphys-033117-053959}.

\bibitem{Montana1992}
C.~Monta{\~{n}}a.
\newblock The colonization of bare areas in two-phase mosaics of an arid
  ecosystem.
\newblock {\em J. Ecol.}, 80(2):315--327, 1992.
\newblock \doi{10.2307/2261014}.

\bibitem{Montana1990}
C.~Monta{\~{n}}a, J.~Lopez-Portillo, and A.~Mauchamp.
\newblock The response of two woody species to the conditions created by a
  shifting ecotone in an arid ecosystem.
\newblock {\em J. Ecol.}, 78(3):789--798, 1990.
\newblock \doi{10.2307/2260899}.

\bibitem{Mueller2013}
J.~Müller.
\newblock Floristic and structural pattern and current distribution of tiger
  bush vegetation in {B}urkina {F}aso ({W}est {A}frica), assessed by means of
  belt transects and spatial analysis.
\newblock {\em Appl. Ecol. Environ. Res.}, 11:153--171, 2013.
\newblock \doi{10.15666/aeer/1102_153171}.

\bibitem{Nathan2013}
J.~Nathan, J.~von Hardenberg, and E.~Meron.
\newblock Spatial instabilities untie the exclusion-principle constraint on
  species coexistence.
\newblock {\em J. Theor. Biol.}, 335:198--204, 2013.
\newblock \doi{10.1016/j.jtbi.2013.06.026}.

\bibitem{Pelletier2012}
J.~D. Pelletier, S.~B. DeLong, C.~A. Orem, P.~Becerra, K.~Compton, K.~Gressett,
  J.~Lyons‐Baral, L.~A. McGuire, J.~L. Molaro, and J.~C. Spinler.
\newblock How do vegetation bands form in dry lands? {I}nsights from numerical
  modeling and field studies in southern {N}evada, {USA}.
\newblock {\em J. Geophys. Res., F: Earth Surface}, 117:F04026, 2012.
\newblock \doi{10.1029/2012JF002465}.

\bibitem{Penny2013}
G.~G. Penny, K.~E. Daniels, and S.~E. Thompson.
\newblock Local properties of patterned vegetation: quantifying endogenous and
  exogenous effects.
\newblock {\em Philos. Trans. R. Soc. London, Ser. A}, 371:20120359, 2013.
\newblock \doi{10.1098/rsta.2012.0359}.

\bibitem{Potapov2005}
A.~B. Potapov and T.~Hillen.
\newblock Metastability in chemotaxis models.
\newblock {\em Journal of Dynamics and Differential Equations}, 17(2):293--330,
  2005.
\newblock \doi{10.1007/s10884-005-2938-3}.

\bibitem{Pueyo2008}
Y.~Pueyo, S.~Kéfi, C.~L. Alados, and M.~Rietkerk.
\newblock Dispersal strategies and spatial organization of vegetation in arid
  ecosystems.
\newblock {\em Oikos}, 117(10):1522--1532, 2008.
\newblock \doi{10.1111/j.0030-1299.2008.16735.x}.

\bibitem{Pueyo2010}
Y.~Pueyo, S.~Kéfi, R.~Díaz-Sierra, C.~Alados, and M.~Rietkerk.
\newblock The role of reproductive plant traits and biotic interactions in the
  dynamics of semi-arid plant communities.
\newblock {\em Theor. Popul. Biol.}, 78(4):289 -- 297, 2010.
\newblock \doi{10.1016/j.tpb.2010.09.001}.

\bibitem{Reynolds2007}
J.~F. Reynolds, D.~M.~S. Smith, E.~F. Lambin, B.~L. Turner, M.~Mortimore,
  S.~P.~J. Batterbury, T.~E. Downing, H.~Dowlatabadi, R.~J. Fernandez, J.~E.
  Herrick, E.~Huber-Sannwald, H.~Jiang, R.~Leemans, T.~Lynam, F.~T. Maestre,
  M.~Ayarza, and B.~Walker.
\newblock Global desertification: Building a science for dryland development.
\newblock {\em Science}, 316(5826):847--851, 2007.
\newblock \doi{10.1126/science.1131634}.

\bibitem{Rietkerk2002}
M.~Rietkerk, M.~C. Boerlijst, F.~{van Langevelde}, R.~HilleRisLambers, J.~{van
  de Koppel}, L.~Kumar, H.~H.~T. Prins, and A.~M. {de Roos}.
\newblock Self‐organization of vegetation in arid ecosystems.
\newblock {\em Am. Nat.}, 160(4):524--530, 2002.
\newblock \doi{10.1086/342078}.

\bibitem{Rietkerk2004}
M.~Rietkerk, S.~C. Dekker, P.~C. {de Ruiter}, and J.~{van de Koppel}.
\newblock Self-organized patchiness and catastrophic shifts in ecosystems.
\newblock {\em Science}, 305(5692):1926--1929, 2004.
\newblock \doi{10.1126/science.1101867}.

\bibitem{Rietkerk2000}
M.~Rietkerk, P.~Ketner, J.~Burger, B.~Hoorens, and H.~Olff.
\newblock Multiscale soil and vegetation patchiness along a gradient of
  herbivore impact in a semi-arid grazing system in {W}est {A}frica.
\newblock {\em Plant Ecol.}, 148(2):207--224, 2000.
\newblock \doi{10.1023/A:1009828432690}.

\bibitem{Rietkerk2008}
M.~Rietkerk and J.~{van de Koppel}.
\newblock Regular pattern formation in real ecosystems.
\newblock {\em Trends Ecol. Evol.}, 23(3):169 -- 175, 2008.
\newblock \doi{10.1016/j.tree.2007.10.013}.

\bibitem{Rodriguez-Iturbe1999}
I.~Rodriguez-Iturbe, A.~Porporato, L.~Ridolfi, V.~Isham, and D.~R. Coxi.
\newblock Probabilistic modelling of water balance at a point: the role of
  climate, soil and vegetation.
\newblock {\em Proc. R. Soc. Lond. A}, 455(1990):3789--3805, 1999.
\newblock \doi{10.1098/rspa.1999.0477}.

\bibitem{Saco2018}
P.~M. Saco, M.~{Moreno-de las Heras}, S.~Keesstra, J.~Baartman, O.~Yetemen, and
  J.~F. Rodriguez.
\newblock Vegetation and soil degradation in drylands: Non linear feedbacks and
  early warning signals.
\newblock {\em Current Opinion in Environmental Science {\&} Health}, 5:67--72,
  2018.
\newblock \doi{10.1016/j.coesh.2018.06.001}.

\bibitem{Salvucci2001}
G.~D. Salvucci.
\newblock Estimating the moisture dependence of root zone water loss using
  conditionally averaged precipitation.
\newblock {\em Water Resour. Res.}, 37(5):1357--1365, 2001.
\newblock \doi{10.1029/2000WR900336}.

\bibitem{Scheiter2007}
S.~Scheiter, S.~Higgins, A.~E. F.~J. Weissing, and E.~M.~A. Geber.
\newblock Partitioning of root and shoot competition and the stability of
  savannas.
\newblock {\em Am. Nat.}, 170(4):587--601, 2007.
\newblock \doi{10.1086/521317}.

\bibitem{Seghieri1997}
J.~Seghieri, S.~Galle, J.~Rajot, and M.~Ehrmann.
\newblock Relationships between soil moisture and growth of herbaceous plants
  in a natural vegetation mosaic in {N}iger.
\newblock {\em J. Arid. Environ.}, 36(1):87--102, 1997.
\newblock \doi{10.1006/jare.1996.0195}.

\bibitem{Serra-Diaz2018}
J.~M. Serra-Diaz, C.~Maxwell, M.~S. Lucash, R.~M. Scheller, D.~M. Laflower,
  A.~D. Miller, A.~J. Tepley, H.~E. Epstein, K.~J. Anderson-Teixeira, and J.~R.
  Thompson.
\newblock Disequilibrium of fire-prone forests sets the stage for a rapid
  decline in conifer dominance during the 21st century.
\newblock {\em Sci. Rep.}, 8:6749, 2018.
\newblock \doi{10.1038/s41598-018-24642-2}.

\bibitem{Sheffer2013}
E.~Sheffer, J.~Hardenberg, H.~Yizhaq, M.~Shachak, E.~Meron, and B.~Blasius.
\newblock Emerged or imposed: a theory on the role of physical templates and
  self‐organisation for vegetation patchiness.
\newblock {\em Ecol. Lett.}, 16(2):127--139, 2013.
\newblock \doi{10.1111/ele.12027}.

\bibitem{Sherratt2005}
J.~A. Sherratt.
\newblock An analysis of vegetation stripe formation in semi-arid landscapes.
\newblock {\em J. Math. Biol.}, 51(2):183--197, 2005.
\newblock \doi{10.1007/s00285-005-0319-5}.

\bibitem{Sherratt2010}
J.~A. Sherratt.
\newblock Pattern solutions of the {K}lausmeier model for banded vegetation in
  semi-arid environments {I}.
\newblock {\em Nonlinearity}, 23(10):2657--2675, 2010.
\newblock \doi{10.1088/0951-7715/23/10/016}.

\bibitem{Sherratt2011}
J.~A. Sherratt.
\newblock Pattern solutions of the {K}lausmeier model for banded vegetation in
  semi-arid environments {II}: patterns with the largest possible propagation
  speeds.
\newblock {\em Proc. R. Soc. Lond. A}, 467(2135):3272--3294, 2011.
\newblock \doi{10.1098/rspa.2011.0194}.

\bibitem{Sherratt2013}
J.~A. Sherratt.
\newblock History-dependent patterns of whole ecosystems.
\newblock {\em Ecol. Complexity}, 14:8--20, 2013.
\newblock \doi{10.1016/j.ecocom.2012.12.002}.

\bibitem{Sherratt2013III}
J.~A. Sherratt.
\newblock Pattern solutions of the {K}lausmeier model for banded vegetation in
  semi-arid environments {III}: The transition between homoclinic solutions.
\newblock {\em Physica D}, 242(1):30 -- 41, 2013.
\newblock \doi{10.1016/j.physd.2012.08.014}.

\bibitem{Sherratt2013IV}
J.~A. Sherratt.
\newblock Pattern solutions of the {K}lausmeier model for banded vegetation in
  semiarid environments {IV}: Slowly moving patterns and their stability.
\newblock {\em SIAM J. Appl. Math.}, 73(1):330--350, 2013.
\newblock \doi{10.1137/120862648}.

\bibitem{Sherratt2013V}
J.~A. Sherratt.
\newblock Pattern solutions of the {K}lausmeier model for banded vegetation in
  semiarid environments {V}: The transition from patterns to desert.
\newblock {\em SIAM J. Appl. Math.}, 73(4):1347--1367, 2013.
\newblock \doi{10.1137/120899510}.

\bibitem{Sherratt2007}
J.~A. Sherratt and G.~J. Lord.
\newblock Nonlinear dynamics and pattern bifurcations in a model for vegetation
  stripes in semi-arid environments.
\newblock {\em Theor. Popul. Biol.}, 71(1):1--11, 2007.
\newblock \doi{10.1016/j.tpb.2006.07.009}.

\bibitem{Siero2018}
E.~Siero.
\newblock Nonlocal grazing in patterned ecosystems.
\newblock {\em J. Theor. Biol.}, 436:64--71, 2018.
\newblock \doi{10.1016/j.jtbi.2017.10.001}.

\bibitem{Siero2019}
E.~Siero, K.~Siteur, A.~Doelman, J.~{van de Koppel}, M.~Rietkerk, and M.~B.
  Eppinga.
\newblock Grazing away the resilience of patterned ecosystems.
\newblock {\em Am. Nat.}, 193(3):472--480, 2019.
\newblock \doi{10.1086/701669}.

\bibitem{Siteur2014a}
K.~Siteur, M.~B. Eppinga, D.~Karssenberg, M.~Baudena, M.~F. Bierkens, and
  M.~Rietkerk.
\newblock How will increases in rainfall intensity affect semiarid ecosystems?
\newblock {\em Water Resour. Res.}, 50(7):5980--6001, 2014.
\newblock \doi{10.1002/2013wr014955}.

\bibitem{Siteur2014}
K.~Siteur, E.~Siero, M.~B. Eppinga, J.~D. Rademacher, A.~Doelman, and
  M.~Rietkerk.
\newblock Beyond {T}uring: The response of patterned ecosystems to
  environmental change.
\newblock {\em Ecol. Complexity}, 20:81 -- 96, 2014.
\newblock \doi{10.1016/j.ecocom.2014.09.002}.

\bibitem{Sprugel1991}
D.~G. Sprugel.
\newblock Disturbance, equilibrium, and environmental variability: What is
  `natural' vegetation in a changing environment?
\newblock {\em Biol. Conserv.}, 58(1):1--18, 1991.
\newblock \doi{10.1016/0006-3207(91)90041-7}.

\bibitem{Svenning2013}
J.-C. Svenning and B.~Sandel.
\newblock Disequilibrium vegetation dynamics under future climate change.
\newblock {\em Am. J. Bot.}, 100(7):1266--1286, 2013.
\newblock \doi{10.3732/ajb.1200469}.

\bibitem{Synodinos2015}
A.~D. Synodinos, B.~Tietjen, and F.~Jeltsch.
\newblock Facilitation in drylands: Modeling a neglected driver of savanna
  dynamics.
\newblock {\em Ecol. Modell.}, 304:11 -- 21, 2015.
\newblock \doi{10.1016/j.ecolmodel.2015.02.015}.

\bibitem{Thiery1995}
J.~M. Thiery, J.-M. D'Herbès, and C.~Valentin.
\newblock A model simulating the genesis of banded vegetation patterns in
  {N}iger.
\newblock {\em J. Ecol.}, 83(3):497--507, 1995.
\newblock \doi{10.2307/2261602}.

\bibitem{Thompson2010}
S.~E. Thompson, C.~J. Harman, P.~Heine, and G.~G. Katul.
\newblock Vegetation‐infiltration relationships across climatic and soil type
  gradients.
\newblock {\em J. Geophys. Res., G: Biogeosci.}, 115:G02023, 2010.
\newblock \doi{10.1029/2009JG001134}.

\bibitem{Tilman1982}
D.~Tilman.
\newblock {\em Resource Competition and Community Structure}.
\newblock Princton University Press, 1982.

\bibitem{Tongway1990}
D.~J. Tongway and J.~A. Ludwig.
\newblock Vegetation and soil patterning in semi-arid mulga lands of {E}astern
  {A}ustralia.
\newblock {\em Aust. J. Ecol.}, 15(1):23--34, 1990.
\newblock \doi{10.1111/j.1442-9993.1990.tb01017.x}.

\bibitem{Tzuk2019}
O.~Tzuk, S.~R. Ujjwal, C.~Fernandez-Oto, M.~Seifan, and E.~Meron.
\newblock Interplay between exogenous and endogenous factors in seasonal
  vegetation oscillations.
\newblock {\em Sci. Rep.}, 9:354, 2019.
\newblock \doi{10.1038/s41598-018-36898-9}.

\bibitem{UNGlobalLandOutlook2017}
{United Nations Convention to Combat Desertification}.
\newblock The global land outlook, 2017.

\bibitem{UNLivestockBrief2005}
{United Nations Food and Agriculture Organization}.
\newblock Livestock sector briefs, 2005.

\bibitem{Ursino2016}
N.~Ursino and C.~Callegaro.
\newblock Diversity without complementarity threatens vegetation patterns in
  arid lands.
\newblock {\em Ecohydrology}, 9(7):1187--1195, 2016.
\newblock \doi{10.1002/eco.1717}.

\bibitem{Ursino2006}
N.~Ursino and S.~Contarini.
\newblock Stability of banded vegetation patterns under seasonal rainfall and
  limited soil moisture storage capacity.
\newblock {\em Adv. Water Resour.}, 29(10):1556 -- 1564, 2006.
\newblock \doi{10.1016/j.advwatres.2005.11.006}.

\bibitem{Valentin1999}
C.~Valentin, J.~d'Herbès, and J.~Poesen.
\newblock Soil and water components of banded vegetation patterns.
\newblock {\em CATENA}, 37(1–2):1--24, 1999.
\newblock \doi{10.1016/S0341-8162(99)00053-3}.

\bibitem{Stelt2013}
S.~{van~der Stelt}, A.~Doelman, G.~Hek, and J.~D.~M. Rademacher.
\newblock Rise and fall of periodic patterns for a generalized
  {K}lausmeier{\textendash}{G}ray{\textendash}{S}cott model.
\newblock {\em J. Nonlinear. Sci.}, 23(1):39--95, 2013.
\newblock \doi{10.1007/s00332-012-9139-0}.

\bibitem{White1971}
L.~P. White.
\newblock Vegetation stripes on sheet wash surfaces.
\newblock {\em J. Ecol.}, 59(2):615--622, 1971.
\newblock \doi{10.2307/2258335}.

\bibitem{Worrall1959}
G.~A. Worrall.
\newblock The {B}utana grass patterns.
\newblock {\em J. Soil Sci.}, 10(1):34--53, 1959.
\newblock \doi{10.1111/j.1365-2389.1959.tb00664.x}.

\bibitem{Zelnik2018}
Y.~R. Zelnik, P.~Gandhi, E.~Knobloch, and E.~Meron.
\newblock Implications of tristability in pattern-forming ecosystems.
\newblock {\em Chaos: An Interdisciplinary Journal of Nonlinear Science},
  28(3):033609, 2018.
\newblock \doi{10.1063/1.5018925}.

\bibitem{Zelnik2013}
Y.~R. Zelnik, S.~Kinast, H.~Yizhaq, G.~Bel, and E.~Meron.
\newblock Regime shifts in models of dryland vegetation.
\newblock {\em Philos. Trans. R. Soc. London, Ser. A}, 371(2004):20120358,
  2013.
\newblock \doi{10.1098/rsta.2012.0358}.

\bibitem{Zimmerman2010}
J.~K. Zimmerman, L.~S. Comita, J.~Thompson, M.~Uriarte, and N.~Brokaw.
\newblock Patch dynamics and community metastability of a subtropical forest:
  compound effects of natural disturbance and human land use.
\newblock {\em Landscape Ecol.}, 25(7):1099--1111, 2010.
\newblock \doi{10.1007/s10980-010-9486-x}.

\end{thebibliography}
\end{document}